\documentclass[fleqn,usenatbib,useAMS]{mnras}

\DeclareRobustCommand{\VAN}[3]{#2}
\let\VANthebibliography\thebibliography
\def\thebibliography{\DeclareRobustCommand{\VAN}[3]{##3}\VANthebibliography}

\usepackage{graphicx}	
\usepackage{amsmath}	
\usepackage{amssymb}	
\usepackage{xcolor,ulem}
\usepackage{comment}
\usepackage{epstopdf}
\usepackage{newtxtext,newtxmath}
\usepackage[T1]{fontenc}

\newcommand{\vhat}{v^{\rm 1D}}
\newcommand{\vbo}{{v}_{\rm bo}^{\rm 1D}}
\newcommand{\tbo}{{t}_{\rm bo}^{\rm 1D}}
\newcommand{\Rbo}{{R}_{\rm bo}^{\rm 1D}}
\newcommand{\ybo}{{y}_{\rm bo}^{\rm 1D}}
\newcommand{\Ebo}{{E}_{\rm bo}^{\rm 1D}}
\newcommand{\Lbo}{{L}_{\rm bo}^{\rm 1D}}
\newcommand{\vstar}{{v}_{*}^{\rm 1D}}
\newcommand{\tstar}{{t}_{*}^{\rm 1D}}
\newcommand{\ts}{{t}_{s}^{\rm 1D}}
\newcommand{\twrap}{{t}_{\rm wrap}}

\title[Light curves of aspherical shock breakout]{Bolometric light curves of aspherical shock breakout}

\author[Irwin, Linial, Nakar, Piran and Sari]{
Christopher M. Irwin,$^{1}$ \thanks{E-mail: christopheri@mail.tau.ac.il, itai.linial@mail.huji.ac.il}
Itai Linial,$^{2}$ \footnotemark[\value{footnote}]
Ehud Nakar,$^{1}$
Tsvi Piran$^{2}$
and Re'em Sari$^{2}$
\\
$^{1}$School of Physics and Astronomy, Tel Aviv University, Tel Aviv 69978, Israel \\
$^{2}$Racah Institute of Physics, The Hebrew University, Jerusalem 91904, Israel}

\date{Accepted XXX. Received YYY; in original form ZZZ}

\pubyear{2021}

\begin{document}
\label{firstpage}
\pagerange{\pageref{firstpage}--\pageref{lastpage}}
\maketitle

\begin{abstract}
The shock breakout emission is the first light that emerges from a supernova. In the spherical case it is characterized by a brief UV flash.
In an axisymmetric, non-spherical prolate explosion,
the shock first breaches the surface along the symmetry axis, then peels around to larger angles, producing a breakout light curve which may differ substantially from the spherically symmetric case.  We study the emergence of a non-relativistic, bipolar shock from a spherical star, and estimate the basic properties of the associated bolometric shock breakout signal.  We identify four possible classes of breakout light curves, depending on the degree of asphericity.  Compared to spherical breakouts, we find that the main distinguishing features of significantly aspherical breakouts are 1) a longer and fainter initial breakout flash and 2) an extended phase of slowly-declining, or even rising, emission which is produced as ejecta flung out by the oblique breakout expand and cool. We find that the breakout flash has a maximum duration of roughly $\sim R_*/v_{\rm bo}$ where $R_*$ is the stellar radius and $v_{\rm bo}$ is the velocity of the fastest-moving ejecta.  For a standard Wolf--Rayet progenitor, the duration of the X-ray flash seen in SN 2008D exceeds this limit, and the same holds true for the prompt X-ray emission of low-luminosity GRBs such as GRB 060218. This suggests that these events cannot be explained by an aspherical explosion within a typical Wolf--Rayet star, implying that they originate from non-standard progenitors with larger breakout radii.

\end{abstract}


\section{Introduction}
\label{sec:intro}

The first light detectable from a stellar explosion is the shock breakout.  Once a radiation-mediated shock with speed $v$ reaches an optical depth of $\sim c/v$ from the edge of the star, where $c$ is the speed of light, the time-scale for radiation to diffuse from the shock to the stellar surface becomes comparable to the time for the shock to cross the same distance, and consequently the radiation from the shock escapes to infinity \citep[see, e.g., the recent review by][and references therein]{levinson20}.  First predicted in the 1970s \citep{colgate74,weaver76,falk78,klein78}, the first detection of soft X-ray emission possibly associated with shock breakout did not come until the late 2000s, when \textit{Swift} observed a bright X-ray flash accompanying the supernova SN 2008D \citep[e.g.,][]{soderberg08,mazzali08,modjaz09}.  Importantly, the breakout light curve provides a way to constrain the stellar radius, which is otherwise difficult to determine when pre-explosion imaging is unavailable. It also provides crucial information about the density structure in the outer layers of massive stars, which in turn may reveal stellar mass-loss during the final stages of a star's life. Observing supernova shock breakout emission to constrain the properties of massive stars is a primary goal of the ULTRASAT mission, which is set to launch in 2024. The potential future mission SIBEX also intends to target shock breakout events, indicating growing observational support for this science \citep{roming21}.  In order to make predictions for future observations by ULTRASAT, it is critically important to improve our theoretical understanding of shock breakout emission, particularly in the case of aspherical shocks where our current  understanding is lacking.

Spherically symmetric shock breakout is by now rather well understood, with a rich history in the literature.  Early work by \citet{gandelman56} and \citet{sakurai60} established a self-similar solution for shock wave propagation in a planar stellar atmosphere with a density which is a power-law in the distance from the stellar surface. \citet{matzner99} expanded on this work to describe the hydrodynamics of the blast wave throughout the entire star, as it traverses the stellar interior, accelerates through the stellar atmosphere, and ultimately breaks out.  Later, these results were extended to cover the trans-relativistic \citep{tan01} and ultra-relativistic \citep{johnson71,pan06,nakar12} regimes.  Many papers have sought to construct analytical light curves at optical-UV frequencies which can be readily compared with observations to extract information on the explosion and progenitor properties \citep[e.g.,][]{chevalier92,chevalier08,nakar10,nakar12,piro10,rabinak11,rabinak12,katz12}.

However, the success of spherical shock breakout theory comes with a caveat: many stellar explosions are likely to be aspherical to some extent.  This is especially true of explosions which take place in rapidly rotating stars and/or are 
accompanied by jets, but even in an initially spherical configuration, mechanisms such as the standing accretion shock instability (SASI) can lead to a degree of inherent asphericity \citep[e.g.,][]{blondin03}. 
In comparison with the spherical case, the breakout of aspherical shocks is much less understood, although significant progress has been made in recent years on both the analytical and numerical fronts.  \citet{matzner13} laid the analytical groundwork for oblique shock breakout, describing the conditions under which obliquity affects the breakout signal, and suggesting the possibility of transient emission produced as ejecta thrown sideways by the oblique shock dissipate their energy outside the star.  In a follow-up paper, \citet{salbi14} investigated the behaviour of the oblique shock near the point of emergence through hydrodynamics simulations.  Later, \citet{afsariardchi18} conducted global simulations investigating the propagation of a non-relativistic, axisymmetric shock through a simple polytropic star, and estimated the frequency-dependent light curve for several progenitor types. In addition, \citet{suzuki16} performed coupled radiation-hydrodynamics simulations of aspherical shock breakout and computed bolometric light curves for a typical blue supergiant progenitor.   On the analytical side, \citet{linial19} recently obtained a solution for the emergence of an oblique shock from a planar surface, which describes the local flow at the point of breakout in terms of the angle between the shock and the surface.  All these works share one important conclusion. The sharp acceleration of the shock as it nears the edge of the star significantly amplifies the shock obliquity. Therefore, even small deviations from sphericity of the shock in the stellar interior can result in shocks which become considerably oblique by the time of breakout. Naturally, this can substantially alter the properties of the breakout signal.

It seems reasonable to expect that many explosions lacking spherical symmetry may retain symmetry about the rotation axis, at least if the cause of the asymmetry is stellar rotation or the formation of jets.  Accordingly, the aim of this paper is to synthesize existing work on spherical and aspherical breakout into a cohesive analytical theory describing how the breakout proceeds in the case of axisymmetry.  To this end, we first develop a simple parametrization for the explosion asymmetry that applies broadly to axisymmetric explosions in general.  Then, for a given parametrization of the asymmetry, we show how to construct an approximate bolometric light curve (assuming the explosion energy and progenitor properties are known), and investigate how the properties of the breakout emission depend on the parameters which control the asphericity.

For typical stellar rotation speeds, the deviation from spherical symmetry is expected to be small.  Furthermore, even if the asymmetry is caused by a jet,  the asphericity at breakout may still be small, if the jet is choked deep enough in the stellar envelope to allow sufficient time for the outflow to spherize before breakout \citep[see, e.g., ][]{irwin19}.  Therefore, in this paper we mainly focus on the case of small asphericity, where the difference between the shock radius and the stellar radius at the onset of breakout is much less than the stellar radius.  The case of more severe asphericity will be left for future work, although we give a brief, qualitative overview of the behaviour in this case in Section~\ref{sec:breakoutclasses}. In addition, in this paper we derive only bolometric light curves, and reserve treatment of the breakout's spectrum for a subsequent paper.

The structure of the paper is as follows.  In Section~\ref{sec:velocityscales}, we review some important previous results on spherical and aspherical shock breakout, and cast them in a new light by considering how the breakout behaviour depends on the inclination angle of the shock when it begins to accelerate. We first consider the structure of the progenitor star in Section~\ref{sec:progenitor}, then review the expected behaviour of a spherical breakout in Section~\ref{sec:spherical}, before addressing the key differences in the non-spherical case in Section~\ref{sec:aspherical}. Next, in Section~\ref{sec:breakoutclasses}, we discuss the expected breakout signal from a generic bipolar shock, and delineate four different breakout scenarios which depend on the extent to which the system is aspherical.  In subsections~\ref{sec:instantaneous} through~\ref{sec:quasioblique}, we discuss three of these scenarios and derive the resulting light curve (the fourth scenario is more complicated and is reserved for future work).  Modifications of the light curve due to light travel time and viewing angle effects, and other caveats, are treated in Section~\ref{sec:lighttravel}. In Section \ref{sec:implications} we discuss astrophysical and observational implications, and we summarize our results and conclusions in Section~\ref{sec:summary}. In addition, to help keep track of the various physical quantities introduced throughout the paper, a table of important symbols and their definitions is provided in Appendix~\ref{appendixa}.

\section{Aspherical breakout: velocity scales and shock inclination}
\label{sec:velocityscales}

A typical star has a density profile which declines sharply in radius near the stellar edge.  When the shock produced by an explosion encounters this steep density gradient, it begins to accelerate, ultimately increasing its speed by a factor of $\sim$ a few to $\sim$ a few tens before the radiation trapped in the shock can diffuse out.  The passage of the shock through the outer parts of the star accelerates a small fraction of the stellar matter to velocities which are significantly faster than the bulk of the supernova ejecta.  The standard shock breakout emission in the spherical theory is comprised of the initial burst of radiation generated as the radiation-mediated shock breaches the stellar surface, and the subsequent cooling emission from the fast shock-accelerated ejecta. Although this fast-moving ejecta contains only a small fraction of the total explosion energy, because of its relatively short emission time-scale and hot temperature, it none the less dominates the supernova emission in UV/X-rays at the earliest times.

Shock acceleration therefore plays a crucial role in shaping the breakout emission, even in the spherical theory, but in the case of an aspherical explosion it is even more important.  The key reason is that the onset of acceleration in the aspherical case is \textit{no longer simultaneous} across the stellar surface. 
Shocked elements which start to accelerate sooner run ahead of ones that accelerate later, causing the shock inclination angle to continuously steepen as the shock nears the surface.  As we show below, this steepening of the shock angle can have profound consequences for the breakout emission, particularly if the shock's inclination angle becomes $\sim$ 1 radian before the shock's radiation can diffuse out (see figure \ref{fig:PlanarVsObliqueBreakout}, and refer to Section~\ref{sec:aspherical} for further discussion).

In this section, we first describe the assumed properties of the progenitor star, and briefly overview some pertinent results and definitions from the 1D spherical breakout theory.  Then, implementing past results on aspherical breakout, we consider the properties of the emerging shock at a given location on the stellar surface (i.e., at a given polar angle $\theta$).  As we will show, key properties of the shock upon breakout (namely, its velocity and inclination angle relative to the surface, and whether it is normal or oblique), ultimately depend only on 1) dimensionless combinations of quantities from the 1D theory; and 2) the shock's inclination angle, $\psi_0(\theta)$, at the radius $R_0$ where the shock begins to accelerate.

Assuming that the opacity and the density structure near the edge of the star have no angular dependence, the quantities from the 1D theory are also independent of $\theta$.  Therefore, for a given progenitor model and explosion energy, the shock breakout behaviour at any point on the stellar surface is set entirely by $\psi_0(\theta)$, which will form the basis of our analysis in Section~\ref{sec:breakoutclasses}.

\subsection{Progenitor structure}
\label{sec:progenitor}

Consider a spherical star with a radius $R_*$ and mass $M_*$.  We assume that in the outer parts of the star, the density takes the form $\rho \propto r^{-\alpha} y^n$, where $y\equiv R_*-r$.  We further suppose that the the transition between the $\rho \propto r^{-\alpha}$ behaviour and the $\rho \propto y^n$ behaviour takes place at a characteristic radius $R_0$, such that we can approximate the density profile as
\begin{eqnarray}
\label{rho}
\rho(r) \approx \begin{cases}
\rho_* \left(\dfrac{r}{R_*}\right)^{-\alpha}, & r \la R_0 \\
\rho_* \left(\dfrac{y}{R_*}\right)^n, & r \ga R_0
\end{cases}.
\end{eqnarray}
The density $\rho_*\equiv \rho(R_0) \sim M_*/R_*^3$, where $M_*$ and $R_*$ are the mass and radius of the star.  The power-law index $n$ has a typical value of $n=3$ for stars with a radiative envelope, or $n=1.5$ for stars with a convective envelope.  The exact value of $\alpha$ is not critical for our analysis, but to give an example, an $n=3$ polytrope has $\alpha \approx 2.5$.

The radius $R_0$ where the density transitions to the steep outer profile is also approximately the radius at which the shock begins to accelerate.  For typical values of $\alpha$ and $n$, $R_0 \simeq R_*/2$.

\subsection{Spherical breakout}
\label{sec:spherical}
Now, suppose that a point explosion with energy $E \gg E_{\rm bind} \approx GM_*^2 /R_*$ is detonated at the centre of the star, causing a strong spherical shock wave to propagate outwards (see however \citet{Linial_Fuller_Sari_21} for a discussion of shock breakout when ${E \ll E_{\rm bind}}$). If $\alpha < 3$ (as is typical), the shock initially decelerates like a Sedov-Taylor blast wave, with $\vhat \propto r^{(\alpha-3)/2}$ for $r<R_0$ \citep{sedov59,taylor50}.   The shock reaches $R_0$ when its age is \citep{sedov59,taylor50}
\begin{eqnarray}
\label{tstar}
\tstar \approx (\rho_* R_*^5/E)^{1/2}
\end{eqnarray} 
and its velocity is 
\begin{eqnarray}
\vstar \approx [2/(5-\alpha)](E/\rho_* R_*^3)^{1/2} \sim (E/M_*)^{1/2}.
\end{eqnarray}
Here, and throughout the paper, we use the superscript `1D' to denote quantities obtained from the 1D theory, which depend only on the explosion energy and the progenitor properties.

Beyond $R_0$, however, the density starts to decline rapidly, and eventually, once ${\rm d}\ln \rho/{\rm d}\ln r < -3$, the shock begins to accelerate. The radius where acceleration sets in differs from $R_0$ only by an order-unity factor depending on $\alpha$ and $n$.  Therefore, we use $R_0$ interchangeably to refer to both the radius where the density breaks to a steeper power-law, and the radius where acceleration begins. The acceleration starts off gradually, but ramps up as the shock approaches the edge of the star.  Near the edge, the velocity scales as $\vhat \propto \rho^{-\mu} \propto y^{-\mu n}$, where $\mu \approx 0.19$ according to the similarity solution of \citet{sakurai60}. See \citet{ro13} for a detailed investigation of the value of $\mu$.

Joining the solutions for $\vhat(r)$ at small and large radii, we obtain
\begin{eqnarray}
\label{v}
\vhat(r) \approx \begin{cases}
\vstar \left(\dfrac{r}{R_*}\right)^{(\alpha-3)/2}, & r \la R_0 \\
\vstar \left(\dfrac{y}{R_*}\right)^{-\mu n}, & r \ga R_0
\end{cases}.
\end{eqnarray}
The optical depth to the shock measured from the stellar edge is $\tau(r) = \int_r^{R_*} \kappa \rho(r') {\rm d}r'$, where $\kappa$ is the opacity, which we take to be constant in the outer layers of the star. For spherical explosions, when $\tau(r)$ becomes equal to ${\approx c/\vhat(r)}$, 
photons in the shock are able to diffuse out, and the shock breaks out.  This occurs when the shock velocity is\footnote{From now on, we neglect for simplicity order--unity factors depending on $\alpha$ and $n$.}
\begin{eqnarray}
\label{vbo}
\vbo \approx \vstar \left( \kappa \rho_* R_0 \dfrac{\vstar}{c} \right)^{\mu n/(1+n-\mu n)},
\end{eqnarray}
at a radius 
\begin{eqnarray}
\Rbo \approx R_*\left[1 - \left(\dfrac{\vbo}{\vstar}\right)^{-1/\mu n}\right] .
\end{eqnarray}
The dimensionless ratio $\vbo/\vstar$ ranges from $\sim 3$ for red supergiants with radius $R_* \sim 500 R_\odot$, to $ \sim 30$ for compact Wolf--Rayet stars with $R_* \sim 5 R_\odot$, with a typical value of $\vbo/\vstar \approx 10$ \citep{nakar10}. Therefore, under typical conditions, $\vbo \gg \vstar$ and  $\Rbo \approx R_*$.
The width of the layer where the flow is affected by photon diffusion, $\ybo \equiv R_*-\Rbo$, therefore satisfies
\begin{eqnarray}
\label{ybo}
\ybo \approx (\vstar/\vbo)^{1/\mu n} R_* \ll R_*.
\end{eqnarray}
This is also the scale over which most of the shock acceleration occurs, i.e. the shock velocity changes by $\sim \vbo$ over a few $\sim \ybo$.  During the breakout, the energy in shells external to $\ybo$,
\begin{eqnarray}
\label{Ebo}
\Ebo \approx 4 \upi R_*^2 \ybo \rho(\ybo) (\vbo)^2 \approx  E \left(\dfrac{\vstar}{\vbo}\right)^{(1+n-2\mu n)/\mu n},
\end{eqnarray}
is released over the diffusion time of the breakout layer (which also equals to its dynamical time),
\begin{eqnarray}
\label{tbo}
\tbo \approx \ybo/\vbo \approx \tstar \left(\dfrac{\vstar}{\vbo}\right)^{(1+\mu n)/\mu n}.
\end{eqnarray}

\begin{figure*}
\centering
\includegraphics[width=\textwidth]{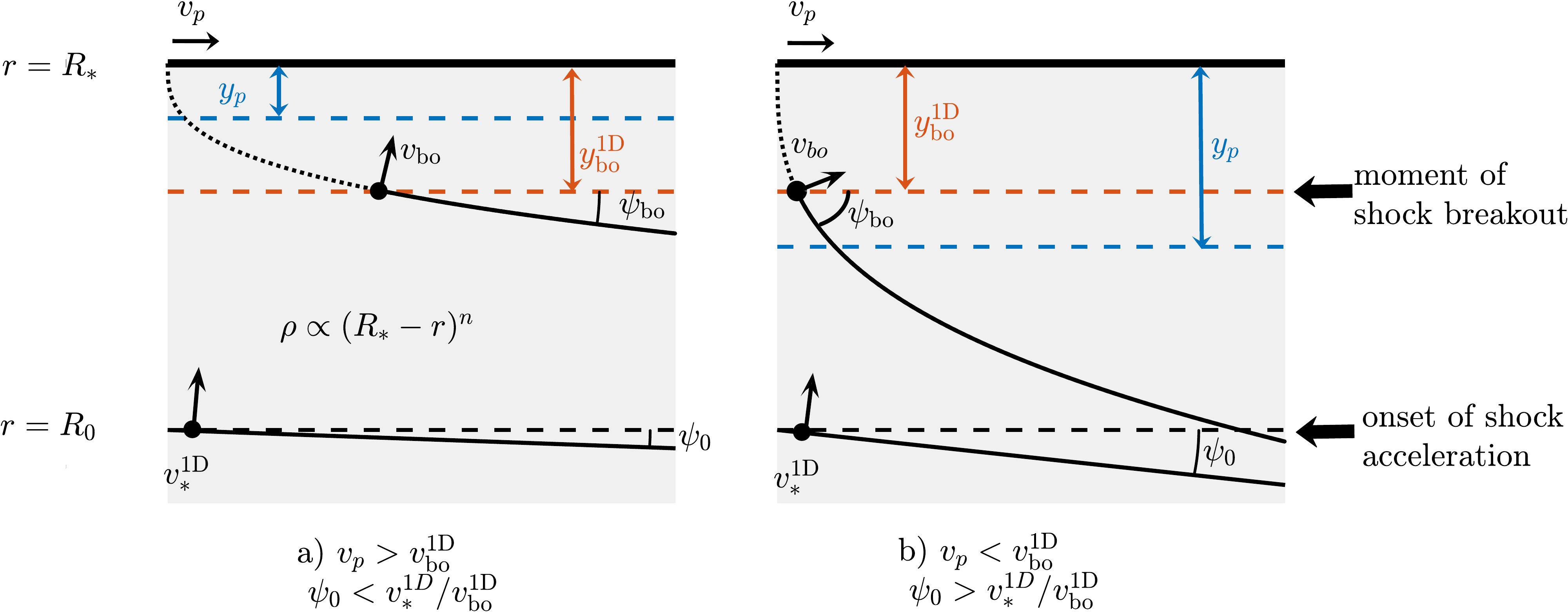}
\caption{Shock propagation near the stellar surface in the plane-parallel and oblique regimes.  The edge of the star where $r=R_*$ is indicated by a thick black line.  The radius $r=R_0$, where the shock crosses into the outer density profile $\rho \propto (R_*-r)^n$ and begins to accelerate, is indicated by a black dashed line.  For each case, the shock is drawn with black lines at two different times: at the onset of shock acceleration, and at the moment of shock breakout.  When the shock begins to accelerate, its speed is $\vstar$ and it is inclined by an angle $\psi_0 \ll 1$ with respect to the radial direction.  The shock steepens as it approaches the stellar edge, eventually breaking out once the optical depth to the shock satisfies $\tau = c/v$, which occurs at a distance of $\ybo$ from the surface.  As the behaviour above $\ybo$ does not affect the emission, the shock there is replaced with a dotted line.  At the moment of breakout, the shock speed is $v_{\rm bo}$ and its inclination angle is $\psi_{\rm bo}$.  The breakout moves along the surface at the pattern speed $v_{\rm p}$.  The breakout properties depend on how $\psi_0$ compares with $\vstar/\vbo$, with two distinct possibilities.  a)~If $\psi_0 \la \vstar/\vbo$, the shock's radiation escapes at $\ybo$ (orange dashed line), \textit{before} the shock becomes oblique at $y_{\rm p}$ (blue dashed line).  In this case the inclination angle upon breakout is $\psi_{\rm bo} \ll 1$ and the breakout is approximately parallel.  The ejecta are accelerated to a maximum velocity of $v_{\rm bo} \approx \vbo$, as in the spherical theory.  The pattern speed $v_{\rm p}\approx \vstar/\psi_0 \approx \vbo/\psi_{\rm bo}$ is greater than $\vbo$.  b)~On the other hand, if $\psi_0 \ga \vstar/\vbo$, then $y_{\rm p} \ga \ybo$ and the radiation cannot diffuse out until \textit{after} the shock becomes highly inclined, so the breakout is oblique.  The inclination angle upon breakout is $\psi_{\rm bo} \sim 1$.  Upon reaching $y_{\rm p}$, the shock is deflected sideways, such that it does not undergo significant acceleration beyond that point.  The maximum ejecta velocity is therefore limited to the shock velocity at $y_{\rm p}$, which is equal to the pattern speed (i.e., $v_{\rm bo} \approx v_{\rm p} \approx \vstar/\psi_0$), and $v_{\rm p} < \vbo$.}
\label{fig:PlanarVsObliqueBreakout}
\end{figure*}

In summary, two characteristic velocity scales appear in the spherical theory: the velocity $\vstar \sim (E/M_*)^{1/2}$ of the shock at $R_0$, just before it enters the layer where the density drops quickly and the shock accelerates, and the final velocity of the shock $\vbo$ obtained when photon diffusion becomes significant and the shock breaks out. Here, for the sake of simplicity, we make the additional assumption that the shock is non-relativistic at breakout, in which case the two velocities satisfy $\vstar \ll \vbo \ll c$.  Note, however, that $\vbo$ can be a few tenths of $c$ for compact stars; in these cases transrelativistic corrections may be required, as discussed by, e.g., \citet{tan01}.

The two characteristic velocities in the problem can be associated with three time-scales: the time ${\tstar \approx R_*/\vstar}$ when the shock reaches the edge of the star, the time $\ts \equiv R_*/\vbo$ when the ejecta with velocity $v_{\rm bo}$ double their radius, and $\tbo$.  In the spherical theory, the duration of observed breakout is $\max(\tbo,R_*/c)$.  The time-scales always satisfy $\tstar \gg \ts \gg \max(\tbo,R_*/c)$; however, both $\tbo < R_*/c$ and $\tbo > R_*/c$ are possible depending on the problem at hand.  In order to simplify the calculations, and facilitate an easier comparison between the spherical and axisymmetric light curves (see Section~\ref{sec:breakoutclasses} for further discussion), we will ignore light travel time effects for the time being, and initially focus on the $\tbo > R_*/c$ case where the initial luminosity of the breakout pulse is set by diffusion through the breakout layer.  In this case, the time-scales satisfy $\tstar \gg \ts \gg \tbo$, and the velocity scales obey $\vstar \ll \vbo \ll R_*/\tbo$. We stress that the $\tbo > R_*/c$ assumption is an idealization. While the ratio $R_*/c\tbo$ is order-unity for red supergiants, it could be as large as $\sim 100$ if the breakout occurs in a Wolf--Rayet star.   Thus, for all but the largest progenitors, $R_*/c > \tbo$ is more typical. We reserve discussion of the more nuanced, but more realistic, case of $R_*/c > \tbo$ to Section~\ref{sec:lighttravel}.

\subsection{Aspherical breakout}
\label{sec:aspherical}
In an aspherical explosion, there are two essential differences. First,
whereas in the 1D case all the quantities related to the breakout are constant across the stellar surface, in the aspherical case the hydrodynamic properties of the breakout vary across the surface.  In the limit where the shock is parallel to the surface, all quantities must converge to their 1D values.

The second difference arises because the shock is no longer parallel to the surface when it breaks out.  Accordingly, at different locations on the stellar surface, the shock breaks out at different times.  This introduces an additional velocity scale: the pattern velocity, $v_{\rm p}$, at which the breakout marches along the surface.  Because $v_{\rm p}$ is a phase velocity, in principle it can exceed $c$.  The spherical case corresponds to the limit $v_{\rm p} \rightarrow \infty$, in which the shock is parallel to the surface. In the other extreme, in which the shock is perpendicular to the surface, the pattern speed is simply equal to the shock velocity just below the surface.

At a given position on the breakout surface, the breakout behaviour depends on how the pattern speed $v_{\rm p}$ compares to the characteristic velocity scales in the 1D problem \citep[as discussed by, e.g.,][]{matzner13}.  There are four possible situations: $v_{\rm p} \ga R_*/\tbo$, ${\vbo \la v_{\rm p} \la R_*/\tbo}$, $\vstar \la v_{\rm p} \la \vbo$, and $v_{\rm p} \sim \vstar$.

The first important distinction with regard to the breakout signal is whether $v_{\rm p}$ is larger or smaller than $\vbo$.  As we will show below, this condition determines whether or not the shock remains a normal shock up until breakout, or becomes an oblique shock.  This distinction is especially relevant to observations, since as we will discuss in Section~\ref{sec:breakoutclasses}, the time-scale and luminosity of the initial breakout are substantially altered for oblique breakouts.  We will refer to the case where the shock remains approximately parallel to the stellar surface up until breakout, so that the breakout signal is not strongly influenced by obliquity, as `plane-parallel breakout,' or simply `parallel breakout.'\footnote{Note that what we call `parallel" breakout has also been referred to as `planar' breakout in the literature.  However, the word `planar' is also commonly used to describe the early dynamical evolution of the post-breakout ejecta where the radius is approximately constant.  To avoid any confusion between the breakout properties and the ejecta properties, we choose to describe the breakout as parallel, and reserve the term planar for descriptions of the ejecta dynamics.}  On the other hand, if the shock becomes substantially oblique before the radiation can escape, we will call this an `oblique breakout.'  The two possible situations (parallel versus oblique breakout) are illustrated in Figure~\ref{fig:PlanarVsObliqueBreakout}.

Now, to determine whether or not the shock becomes oblique before breaking out, let us consider a small patch of the shock surface which is tilted by a small angle $\psi_0$ with respect to the radial direction when it reaches $R_0$, as depicted by the lowermost solid black curves in  Figure~\ref{fig:PlanarVsObliqueBreakout}.  One side of the patch starts to accelerate sooner than the other, causing the shock angle to steepen as it propagates towards the edge.  Under the approximation that the shock motion remains nearly radial, the shock inclination angle at a distance $y$ from the edge of the star is given by
\begin{eqnarray}
\label{psi_y}
\tan \psi (y) \approx (y/R_*)^{-\mu n} \psi_0 .
\end{eqnarray}
Once $\tan \psi (y) \sim 1$, the shock becomes strongly oblique and the approximation of purely radial motion breaks down.  Applying this condition to equation~\ref{psi_y}, we obtain a characteristic length scale, 
\begin{eqnarray}
\label{Rp}
y_{\rm p} \approx \psi_0^{1/\mu n} R_*,
\end{eqnarray}
over which the shock angle becomes highly inclined.  Put another way, non-radial motion is important only down to a depth of $\sim y_{\rm p}$ from the edge of the star. For $\psi_0 \ll 1$, $y_{\rm p} \ll R_*$ and obliquity only affects the flow near the edge of the star.

We define the pattern speed $v_{\rm p}$ as the speed of the breakout as measured along the stellar surface at $r=R_*$.  Note that, in general, $v_{\rm p}$ can be a function of $r$.  However, as we will show below, the pattern speed is approximately constant with radius above $r=R_0$.  Therefore, the pattern speed of the breakout moving along the surface at $r=R_*$ is effectively the same as the pattern speed of the point where the shock crosses the boundary at $r=R_0$ and begins to accelerate; we use $v_{\rm p}$ to refer to both of these quantities interchangeably.  
The pattern speed $v_{\rm p}$ (as measured at $r=R_0$) 
is directly related to the initial shock speed $v_*$ and shock inclination angle $\psi_0$ via 
\begin{eqnarray}
\label{vp}
v_{\rm p} = v_*/\sin \psi_0 \approx \vstar/\psi_0,
\end{eqnarray}
where in the second equality we have assumed that $\psi_0 \ll 1$ and $v_* \approx \vstar$ apply across the whole surface.  (This corresponds to the case where although the shock is not spherical, the total shocked volume is still $\approx \frac{4}{3}\upi R_*^3$; for more discussion about this assumption, see Section~\ref{sec:breakoutclasses}). As expected, we obtain $v_{\rm p} \rightarrow \infty$ for a shock parallel to the surface, and $v_{\rm p} = v_*$ for a perpendicular shock.  Combining expressions~\ref{Rp} and~\ref{vp}, the obliquity length scale in the limit $\psi_0 \ll 1$ can be expressed in terms of the pattern velocity as $y_{\rm p} \approx (\vstar/v_{\rm p})^{1/\mu n} R_*$, reproducing the result previously derived by \citet{matzner13}. Note that at a depth $y=y_{\rm p}$, the shock velocity is $\approx v_{\rm p}$ via equation~\ref{v}.

We can now determine whether the shock is a normal shock or an oblique shock at the moment of breakout by comparing the length scales $\ybo$ and $y_{\rm p}$.  First, if $\psi_0 \la \vstar/\vbo$, then $v_{\rm p} \ga v_{\rm bo}$ and $ \ybo \ga y_{\rm p}$ 
In this case, photons start to diffuse out of the flow before the shock angle steepens significantly, as in panel~a) of Figure~\ref{fig:PlanarVsObliqueBreakout}. The shock velocity at the moment of breakout, $v_{\rm bo}$, is approximately the same as in the spherical case, i.e. $v_{\rm bo}\approx \vbo$. In addition, the shock inclination upon breakout, $\psi_{\rm bo}$, satisfies 
\begin{eqnarray}
\label{psibo}
\psi_{\rm bo} = \psi(\ybo) \approx (\vbo/\vhat_0) \psi_0 \la 1.
\end{eqnarray}
This plane-parallel breakout regime can be further subdivided into the cases $\psi_0 \la \vstar \tbo/R_*$ ($v_{\rm p} \ga R_*/\tbo$) and $\vstar \tbo/R_* \la \psi_0 \la \vstar/\vbo$ ($\vbo \la v_{\rm p} \la R_*/\tbo$), which we refer to as `instantaneous' and `non-instantaneous' plane-parallel breakouts, respectively.  The difference between the two cases is that in the instantaneous case, the time for the breakout to cross the star is short compared to the diffusion time of the breakout layer, so any emission is smeared over a time-scale $\tbo$.  In the non-instantaneous case,  the diffusion time is comparatively short, and the duration of emission is set by the time for the breakout to wrap around the star instead (for further discussion, see Section~\ref{sec:breakoutclasses}).  

If $\psi_0 \ga \vstar/\vbo$, on the other hand, then $v_{\rm p} \la \vbo$ and the shock becomes an oblique shock before breaking out, as in panel b) of Figure~\ref{fig:PlanarVsObliqueBreakout}.  As shown by various authors \citep{matzner13,linial19}, in this case material down to a depth of $\sim y_{\rm p}$ gets deflected sideways. Once the shock motion becomes non-radial, the shock does not undergo significant further acceleration, such that the ejecta velocities are effectively limited to the shock velocity at $y_{\rm p}$, which is $v_{\rm p}$\footnote{The fact that the ejecta velocity is limited to the pattern speed can be intuitively understood by the following argument. In a reference frame in which the breakout point is stationary, cold unshocked material approaches a stationary oblique shock at the pattern speed $v_{\rm p}$. In this frame, the flow is steady, and the Bernoulli equation implies that enthalpy is conserved along streamlines. As the shocked material is being ejected and cools down, its terminal velocity is limited to $v_{\rm p}$. Switching back to the lab frame, the ejecta velocity is limited to $2v_{\rm p}$.}. The breakout velocity in this case therefore satisfies  $v_{\rm bo} \approx v_{\rm p} \la \vbo$, while the inclination angle at breakout is $\psi_{\rm bo} \sim 1$ radian.
As in the plane-parallel breakout case, the oblique breakout regime can be subdivided into two cases.  First, if $\vstar/\vbo \la \psi_0 \ll 1$ ($\vstar \ll v_{\rm p} \la \vbo$), then $R_* \gg y_{\rm p} \ga \ybo$ so that the non-radial motions are confined to a thin layer near the edge of the star.  We refer to this as the `locally oblique' regime.  However, if $\psi_0 \sim 1$ ($v_{\rm p} \sim \vstar$), then the obliquity reaches down to depths of order $R_*$, with $y_{\rm p} \sim R_* \gg \ybo$.  This `globally oblique' limit describes the case in which the shock was already highly inclined when it reached $R_0$.  We caution that the approximation used in equation~\ref{vp}, which assumed $\psi_0 \ll 1$, does not apply in the globally oblique regime.

An interesting corollary to the above results is that approximately the same pattern speed is obtained whether it is computed at $R_0$ where the shock begins to accelerate, or at $\approx R_*$ where the shock ultimately breaks out.  We argue that this is true regardless of the breakout regime.  For the case of a plane-parallel breakout, where the shock velocity remains nearly radial, this occurs because the inclination angle at a depth $y$ is $\psi(y) \approx [v(y)/\vstar] \psi_0 \propto v(y)$ according to equation~\ref{psi_y}. The pattern speed, $v_{\rm p}(y)=v(y)/\sin[(\psi(y)] \approx \vstar/\psi_0$, is therefore approximately independent of depth.  For a locally oblique breakout, the shock angle steepens to $\psi(y) \sim 1$ radian for $y < y_{\rm p}$, but the shock velocity in that region is limited to the velocity at the obliquity depth, $v(y_{\rm p})$.  Thus, the pattern speed of the breakout is $v_{\rm p}(y) = v(y_{\rm p})/\sin[\psi(y)] \approx (\vstar/\psi_0)/(1)$, and we recover the same result as before.   In the globally oblique case, $\psi(y) \sim \psi_0 \sim 1$ everywhere, and the pattern speed is $v_{\rm p} \sim \vstar$ at all radii, so the expression $v_{\rm p} \approx \vstar/\psi_0$ still roughly holds.  Thus, in order to estimate the pattern speed of the breakout at any point on the surface, it is sufficient (to a first approximation) to calculate the pattern speed at the base of the acceleration region (i.e., at $r=R_0$).  This is convenient, because it means that the breakout pattern speed depends mainly on assumptions about the underlying asphericity, and is mostly insensitive to assumptions about the density profile in the outer parts of the star.  We will use this result later on in Section~\ref{sec:breakoutclasses}.

\section{Breakout emission from bipolar shocks: four classes of breakout}
\label{sec:breakoutclasses}

Now that we understand how the characteristics of the breakout at a given point on the surface depend on the initial inclination angle $\psi_0$, let us consider how the breakout behaves across the whole surface of the star for a given initial shock shape described by $\psi_0=\psi_0(\theta)$, where $\theta$ is the polar angle.  Here we consider the case of a bipolar shock, elongated along the symmetry axis, breaking out of a spherical star. 

We adopt an idealized parametrization of $\psi_0(\theta)$ in which the initial inclination angle is a power-law in $\theta$:\footnote{Here, we assume for simplicity that $\psi_0(\theta)$ has the same functional form all the way up to $\theta=\upi/2$, which may not be the case in a realistic explosion (as will be discussed below).}
\begin{eqnarray}
\label{psiansatz}
\psi_0(\theta) = \epsilon (2\theta/\upi)^k,
\end{eqnarray}
where have chosen the dimensionless normalization factor $\epsilon$ such that 
\begin{eqnarray}
\label{epsilondef}
\epsilon \equiv \psi_0(\theta = \upi/2).
\end{eqnarray}
We assume that equation~\ref{psiansatz} holds across the whole surface of the star.  This form for $\psi_0(\theta)$ has the advantage of being easy to work with, while being general enough to describe a wide variety of shock shapes.  More complicated shock shapes can be reduced to the form in equation~\ref{psiansatz} for small $\theta$ by expanding to leading order in $\theta$, so this parametrization should at least be reasonable close to the axis. As an example, for a prolate ellipsoidal shock with eccentricity $e_0$ inside a spherical star, $\psi_0(\theta)$ is described by equation~\ref{psiansatz} with $k=1$ and $\epsilon=\frac{\upi}{2} e_0^2/(1-e_0^2)$ for $\theta \ll \sqrt{1-e_0^2}$. If the eccentricity is small, $e_0 \ll 1$, then equation \ref{psiansatz} is a good approximation up to roughly $\theta \approx \upi/4$, with $\psi_0(\theta) \approx e_0^2 \theta$. The case of a spherical shock, which has $\psi_0=0$ for all $\theta<\upi/2$, corresponds to the limit $k\rightarrow \infty$. 

In order to keep the problem tractable, we consider only the case where $\psi_0$ is monotonically increasing with $\theta$, but it is worth mentioning that under realistic conditions, $\psi_0$ is also expected to go to zero at the equator, which is not captured by equation~\ref{psiansatz}.  The behaviour during the latter parts of the breakout, when the shock nears the equator, may therefore differ somewhat from what we predict, particularly for off-axis observers (as will be discussed in section \ref{sec:lighttravel}). For on-axis observers, our predictions are likely to remain mostly valid as long as $\psi_0$ is increasing up to $\theta \sim 1$ radian, because the region where $\theta \ga 1$ has a relatively small projected area and therefore does not contribute much to the overall emission.  
Furthermore, as we later show, if $\psi_0$ exceeds $\sim \vstar/\vbo$ at latitude $\theta_{\rm obq}$, the breakout ejecta spreads on non-radial trajectories, concealing the stellar surface at higher latitudes $\theta>\theta_{\rm obq}$.
Therefore, as long as equation~\ref{psiansatz} applies up to $\theta_{\rm obq}$, the lightcurve is essentially unchanged from the case we consider, even if $\psi_0$ deviates from the form of equation~\ref{psiansatz} at $\theta> \theta_{\rm obq}$.

\begin{figure*}
\centering
\includegraphics[width=\textwidth]{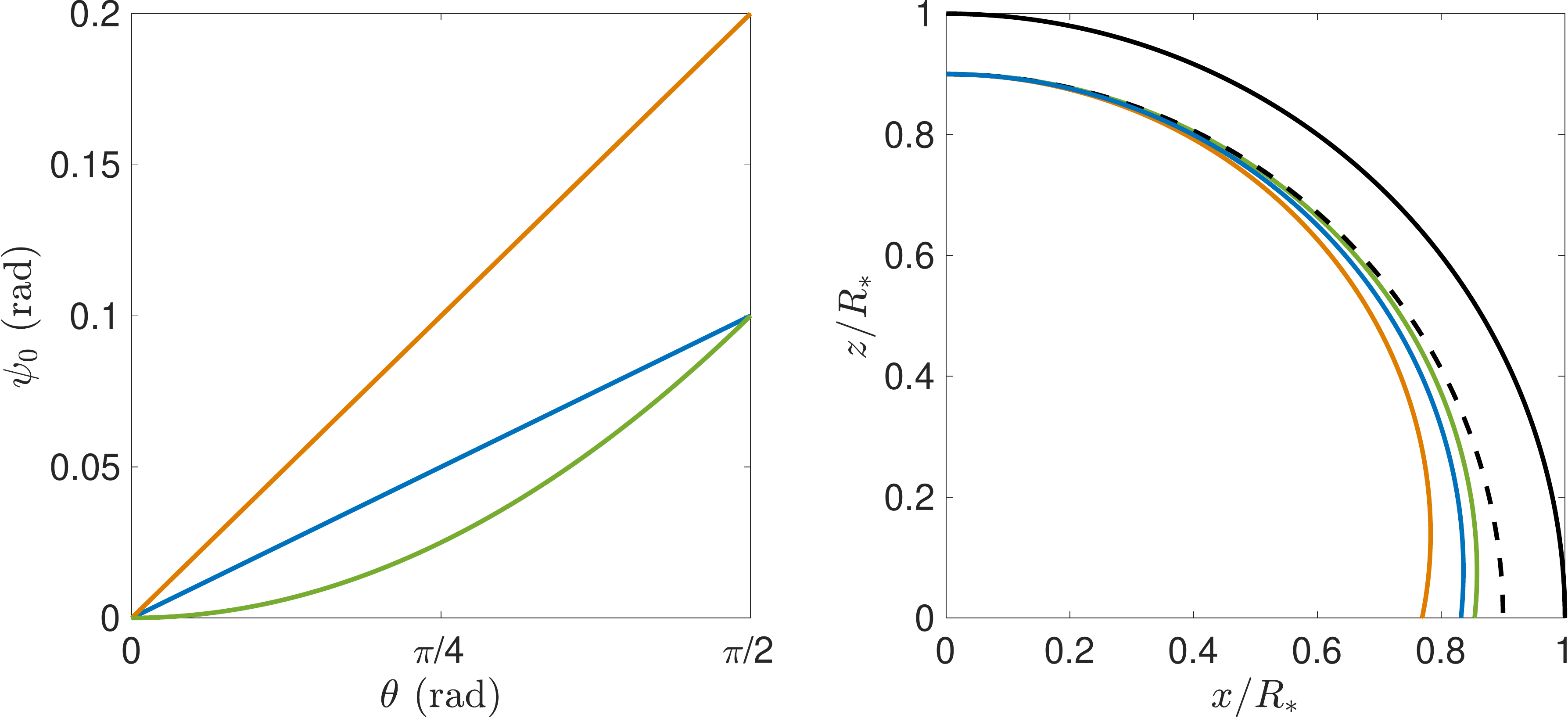}
\caption{Comparison of three different parametrizations for $\psi_0$ using equation~\ref{psiansatz}: $\epsilon=0.1$ and $k=1$ (blue), $\epsilon=0.1$ and $k=2$ (green), and $\epsilon=0.2$ and $k=1$ (orange). \textit{Left:} $\psi_0$ as a function of $\theta$ in each case.  \textit{Right:} The approximate shape of the shock at the onset of shock acceleration in each model.  The stellar surface is indicated by a solid black line.  The radius $R_0$ where shock acceleration begins, which is chosen to be $R_0=0.9R_*$, is denoted by a black dashed line, .}
\label{fig:AsphericalShockParametrization}
\end{figure*}

For small $\psi_0$, equation~\ref{psiansatz} can be converted into an approximate expression for the shock radius, $r_{\rm s}$, as a function of $\theta$.  Over a small angular distance ${\rm d}\theta$, the shock radius changes by an amount ${\rm d}r_{\rm s} = -\psi(\theta,r_{\rm s}) r_{\rm s} \, {\rm d}\theta$, where $\psi(\theta,r_{\rm s})$ is the shock inclination angle at latitude $\theta$, at radius $r_{\rm s}$. Around $r_{\rm s} \approx R_0$, the above relation can be approximated as ${\rm d}r_{\rm s} \approx -\psi_0(\theta) R_0 \, {\rm d}\theta$. Substituting equation \ref{psiansatz} for $\psi_0(\theta)$ we find
\begin{equation}
    r_{\rm s}(\theta) \approx R_0 \left[ 1- \epsilon (2/\upi)^k \theta^{1+k}/(1+k)\right] \,
\end{equation}
to leading order in $\theta$, where we have used the initial condition $r_{\rm s}=R_0$ at $\theta=0$.  In Figure~\ref{fig:AsphericalShockParametrization}, we plot $\psi_0(\theta)$ and compare the resulting shape of the shock at the onset of shock acceleration for three different choices of the parameters $\epsilon$ and $k$.

\begin{figure*}
    \centering
    \includegraphics[width=\textwidth]{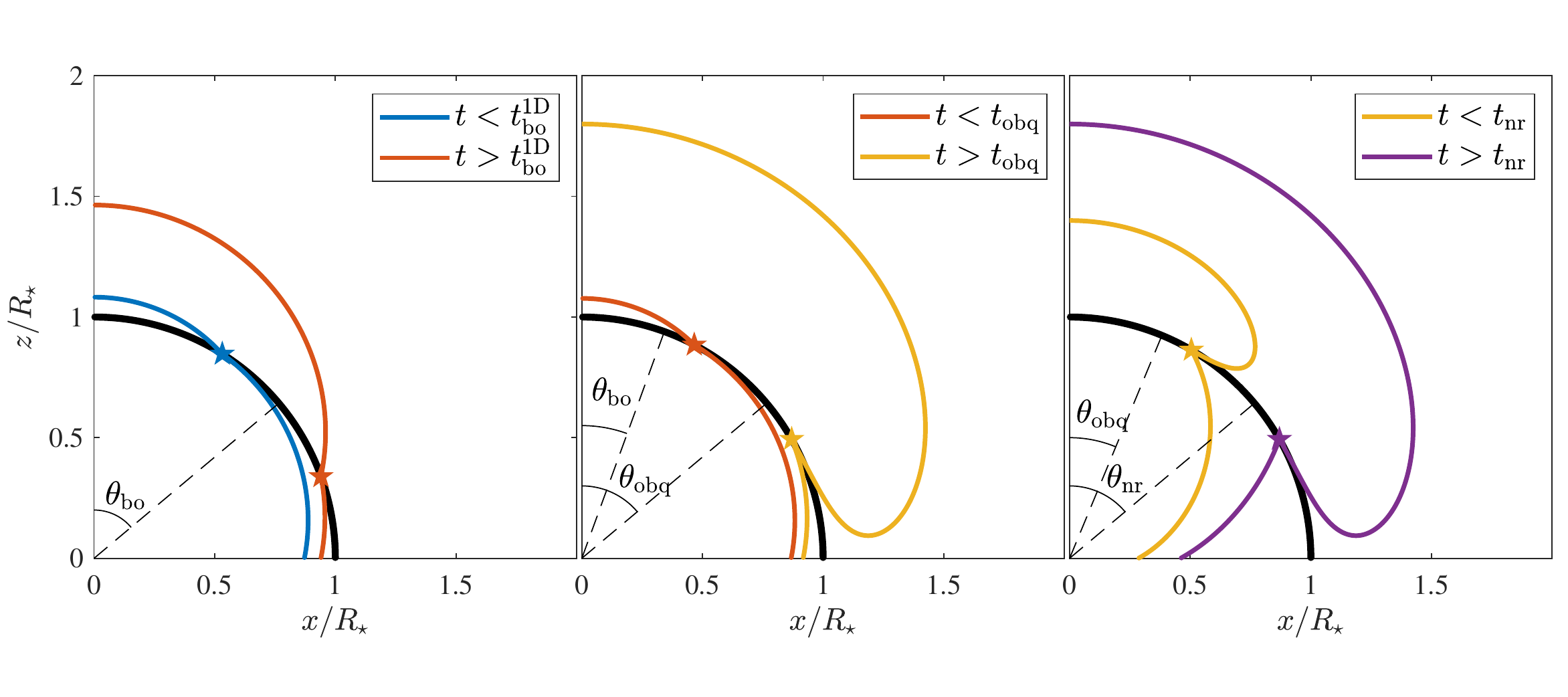}
    \caption{Schematic illustrating three out of the four regimes of aspherical breakout discussed in the paper - non-instantaneous quasi-spherical plane-parallel (left), quasi-spherical oblique (centre) and significantly aspherical (right). Each panel shows the shock front inside the star and the extent of the outermost ejecta at two representative times. A non-instantaneous quasi-spherical plane parallel breakout is characterized by an angle $\theta_{\rm bo}$ at which the shock breaks out at time $\tbo$, marking the transition to times at which the breakout's non-instantaneity becomes apparent. Note that material is ejected radially across the entire stellar surface in this regime.
    In the next regime, the breakout is initially parallel and ejecta follow radial trajectories. However, beyond $\theta_{\rm obq}$ the breakout becomes oblique, and material is thrown sideways to form a sideways-spreading cloud of ejecta, shrouding the breakout point (marked with a star) from most observers. Finally, in the significantly aspherical regime, the shock inclination angle is of order unity even before the shock begins to accelerate in the steep density profile near the stellar edge.}
    \label{fig:SchematicBreakoutClasses}
\end{figure*}

\begin{table*}
\centering
\begin{tabular}{| c | c | c | c | c |}
\hline 
\, & Effectively & Non-instantaneous & Non-instantaneous & Significantly \\
$\psi_0 = \epsilon \theta^k$ & instantaneous & quasi-spherical & quasi-spherical & non-spherical \\
\, & spherical & plane-parallel & oblique & oblique \\
\hline
$\epsilon$ & $\epsilon \la \tbo/\tstar$ & $\tbo/\tstar \la \epsilon \la \vstar/\vbo$ & $\vstar/\vbo \la \epsilon \la 1$ & $\epsilon \ga 1$ \\ \hline
$\twrap$ & $\twrap \la \tbo$ & $\tbo \la \twrap \la \ts$ & $\ts \la \twrap \la \tstar$ & $\twrap \sim \tstar$\\ \hline
$\theta_{\rm bo}$ & - & $(\tbo/\twrap)^{1/(1+k)}$ & $(\tbo/\twrap)^{1/(1+k)}$ & $(\tbo/\twrap)^{1/(1+k)}$\\ \hline
$\theta_{\rm obq}$ & - & - & $(\ts/\twrap)^{1/(1+k)}$ & $(\ts/\twrap)^{1/(1+k)}$\\ \hline
$\theta_{\rm nr}$ & - & - & - & $\epsilon^{-1/k}$\\ \hline 
\end{tabular}
\caption{Summary of key properties for each of the regimes discussed in Section~\ref{sec:breakoutclasses}. Order-unity prefactors are omitted for clarity. }
\label{table:regimes}
\end{table*}

We now use the results of Section~\ref{sec:velocityscales} to compute the pattern speed.  For cases of slight asphericity (i.e., $\epsilon \ll 1$), where the volume contained in the shocked region is roughly equal to the star's volume, the shock's initial velocity when it starts to accelerate, $v_*(\theta)$, is approximately independent of angle: $v_*(\theta) \approx \vstar \sim (E/M_*)^{1/2}$.  The pattern velocity is then given by substituting equation~\ref{psiansatz} into equation~\ref{vp} to obtain
\begin{eqnarray}
\label{vptheta}
v_{\rm p}(\theta) \approx \vstar \epsilon^{-1} (2\theta/\upi)^{-k},
\end{eqnarray}
which is valid for $\epsilon \ll 1$. Note that in this limit, $\epsilon$ is related to the pattern speed at the equator in a simple way: $\epsilon \approx \vstar/v_{\rm p}(\theta=\upi/2)$.  

The time for the breakout to travel a distance $R_* {\rm d}\theta$ along the surface is ${\rm d}t=R_* {\rm d}\theta/v_{\rm p}(\theta)$.  Integrating subject to the initial condition $t=0$ at $\theta=0$, we obtain an expression for the time it takes for the breakout to reach the angle $\theta$:
\begin{eqnarray}
\label{ttheta}
t(\theta) & = & \int_0^{\theta} \dfrac{R_*}{v_{\rm p}(\theta)} {\rm d}\theta
\\
\nonumber
& \approx & \dfrac{\upi}{2}\left(\dfrac{\epsilon \tstar}{1+k}\right) \left(\dfrac{2\theta}{\upi}\right)^{1+k}.
\end{eqnarray}
Therefore, the total time it takes for the breakout to wrap around to the equator is given by
\begin{eqnarray}
\label{twrap} 
\twrap \equiv t(\theta=\upi/2) \approx \dfrac{\upi}{2}\left(\dfrac{\epsilon \tstar}{1+k}\right).
\end{eqnarray}
As expected, when $\epsilon \rightarrow 0$ or $k\rightarrow \infty$, $\twrap \rightarrow 0$ and we return to the limit of an instantaneous spherical breakout.

At each point on the breakout surface, the properties of the breakout depend on the value of $\psi_0$ (as discussed in Section~\ref{sec:velocityscales}), with the breakout being parallel and instantaneous for $\psi_0 \la \vstar \tbo/R_*$, parallel and non-instantaneous for $\vstar \tbo/R_* \la \psi_0 \la \vstar/\vbo$, locally oblique for $\vstar/\vbo \la \psi_0 \la 1$, and globally oblique for $\psi_0 \sim 1$.  In our parametrized model, $\psi_0$ always has a value of 0 at $\theta=0$, and increases to a value of $\min(\epsilon,1)$ at $\theta=\upi/2$.  The fact that $\psi \rightarrow 0$ on the axis guarantees that there will always be a region of plane-parallel breakout at sufficiently small angles.  

Going beyond that, we can identify up to three critical angles at which the breakout behaviour or emission changes.  First, recalling that any signals shorter than $\tbo$ will be smeared by diffusion through the breakout layer, we consider the angular extent $\theta_{\rm bo}$ of the region where this smearing is important.
Setting $t(\theta)=\tbo$ in equation~\ref{ttheta}, we obtain
\begin{equation}
\label{thetabo}
    \theta_{\rm bo} \approx \frac{\upi}{2} \times \min\left\{ \left( \frac{\tbo}{t_*^{\rm 1D}} \frac{1+k}{\epsilon} \right)^{1/(1+k)} , 1 \right\} \,.
\end{equation}
Next, we use the result that the shock becomes oblique when $v_{\rm p} < \vbo$ to calculate the second critical angle, $\theta_{\rm obq}$, at which the breakout transitions from the parallel regime to the oblique regime.  Substituting $v_{\rm p}=\vbo$ into equation~\ref{vptheta} yields
\begin{equation}
    \theta_{\rm obq} \approx \frac{\upi}{2} \times \min\left\{ \left( \frac{v_{*}^{\rm 1D}}{\vbo} \epsilon^{-1} \right)^{1/k} , 1 \right\} \,.
\end{equation}
Finally, we determine the angle $\theta_{\rm nr}$, where the breakout enters the globally oblique regime and the gas velocities become \textit{non-radial} even before shock acceleration by taking $\psi_0(\theta) = 1$ in equation~\ref{psiansatz}, with the result
\begin{equation}
    \theta_{\rm nr} \approx \frac{\upi}{2} \times \min\left(
    \epsilon^{-1/k},1 \right) \,.
\end{equation}

Depending on whether zero, one, two, or all three  of these critical angles are relevant, we can identify four possible breakout scenarios, with the relevant scenario determined by the value of $\epsilon/(1+k)$.  In three of the four scenarios, asphericity significantly alters the breakout emission; these three cases are illustrated in Figure 3.  The four scenarios are as follows: 
\begin{itemize}
\item \textit{Effectively instantaneous spherical breakout:}  If $\epsilon/(1+k) \la \vstar \tbo/R_*$, then $\psi_0 \la \vstar \tbo/R_*$ is satisfied for all $\theta$.  In this situation, the time for the breakout to wrap around the star, $\twrap$, is less than the diffusion time of the breakout layer, $\tbo$.  The breakout emission is smeared over a diffusion time and is effectively indistinguishable from that of a completely spherical breakout.
\item \textit{Non-instantaneous quasi-spherical plane-parallel breakout:} If $\vstar \tbo/R_* \la \epsilon/(1+k) \la \vstar/\vbo$, then $\theta_{\rm bo}<1$.  In this case $\psi_0 \la \vstar \tbo/R_*$ holds at small $\theta$, but $\vstar \tbo/R_* \la \psi_0 \la \vstar/\vbo$ obtains over the rest of the surface.  There is a region near the axis with $\theta<\theta_{\rm bo}$ whose emission is smeared by diffusion (as depicted in the left panel of figure~\ref{fig:SchematicBreakoutClasses}), but the overall time-scale of the breakout flash is set by the time it takes the breakout to reach the equator, $\twrap$.  However, the breakout remains effectively parallel over the whole stellar surface.  In addition, all of the ejecta travel on essentially radial trajectories.
\item \textit{Non-instantaneous quasi-spherical oblique breakout:} If $\vstar/\vbo \la \epsilon/(1+k) \la 1$, then $\theta_{\rm bo}<\theta_{\rm obq}<1$, as shown in the middle panel of Figure~\ref{fig:SchematicBreakoutClasses}. There is region of plane-parallel breakout near the axis, where $\theta<\theta_{\rm obq}$ and $\psi_0 \la \vstar/\vbo$. However, over most of the surface, $\psi_0 \ga \vstar/\vbo$ and therefore the breakout is oblique.  None the less, $\psi_0 \la 1$ still applies at all $\theta$, indicating that the initial shape of the shock is roughly spherical.  Once the breakout becomes oblique, ejecta are thrown sideways from the breakout point, and the resulting mushroom-shaped ejecta cloud hides the breakout from view (as shown by the yellow curve in Figure~\ref{fig:SchematicBreakoutClasses}).  Therefore, the duration of the breakout flash is set by the time for the shock to cross the plane-parallel breakout region, which is $\approx R_* \theta_{\rm obq}/\vbo$. In this case, the velocities in the stellar interior are predominantly radial, but the velocities near the stellar edge (and also outside of the star) have a significant non-radial component.
\item \textit{Significantly aspherical oblique breakout:} Finally, if $\epsilon/(1+k) \ga 1$, then $\theta_{\rm bo}<\theta_{\rm obq}<\theta_{\rm nr}<1$, and the shock shows considerable departures from spherical symmetry even before shock acceleration begins, as depicted in the right panel of Figure~\ref{fig:SchematicBreakoutClasses}.  In other words, when the on-axis material first starts to accelerate, the region enclosed by the shock has an opening angle of $\la 1$ radian.  Although there is still a region near the axis where the breakout is parallel, it is relatively small, so that the breakout is locally or globally oblique over almost the entire breakout surface.  There are substantial non-radial velocities at all radii, both inside and outside of the star. 
\end{itemize}

The relevant values of $\epsilon$, $\twrap$, and the three critical angles $\theta_{\rm bo}$, $\theta_{\rm obq}$, and $\theta_{\rm nr}$ in each regimes are summarized in Table~\ref{table:regimes}.

In the first three scenarios, the difference between the shock radius along the poles and the shock radius at the equator is of order $\sim \epsilon R_*$, which is small compared to $R_*$ since $\epsilon < 1$ in these regimes.  Therefore, the shape of the shocked region does not change appreciably during the breakout, and the volume of the shocked region is approximately constant and equal to $\approx \frac{4}{3}\upi R_*^3$ throughout the entire breakout.  Consequently, the shock speed at the base of the acceleration zone, $v_* \approx \vstar$, is nearly independent of $\theta$, so that the behaviour at each point on the surface is determined only by the outer density profile and the inclination angle $\psi_0(\theta)$.  

The significantly aspherical case, however, is considerably more complicated because both the shape and volume of the shocked region change appreciably over the course of the breakout.  As a result, not only $\psi_0$ but also $v_*$ will vary with $\theta$. In this case, at the time when the shock first breaks out at the poles, the shock at larger $\theta$ is still located deep within the star.  As the breakout proceeds, the ejecta in this high-latitude region follow curved and highly non-radial trajectories towards the surface, decelerating continuously along the way. By the time a given point on the shock reaches $R_0$, both its speed and inclination may have changed considerably from their initial values when the breakout began. 
To treat this case properly, the trajectory of each point along the shock would have to be calculated and used to connect the shock's initial properties with its properties at $R_0$, but this is beyond the scope of this paper. 

Because of these complications, we reserve discussion of the significantly aspherical case for future work, and discuss here only the first three scenarios, for which $\epsilon \la 1$ and equation~\ref{vptheta} applies across the whole stellar surface. 
Below, we discuss each of these scenarios in turn, and present a method for estimating the bolometric light curve in aspherical breakouts.  For simplicity, we assume for now that the observer is located along the axis of the system, and consider only the case where $\tbo > R_*/c$ (as discussed previously in Section 2). After discussing the emission in this idealized situation, we will address the effects of relaxing these assumptions and compute more realistic light curves in Sections~\ref{sec:lighttravel}.

\subsection{Effectively instantaneous spherical breakout}
\label{sec:instantaneous}

The emission following the shock breakout in the case of full spherical symmetry was discussed in detail by, e.g., \citet{nakar10}; here, we simply review the salient results.  Conceptually, it is useful to think of the post-breakout flow as a succession of spherical shells, each with a characteristic mass and velocity.  The first shell to be observed has a velocity $\vbo$ and an initial width $\ybo$, and contains energy $\Ebo$ (as discussed in Section~\ref{sec:velocityscales}).  Following \citet{nakar10}, we refer to this as the `breakout shell.'  Later on, the shell seen by a distant observer is the one which satisfies $\tau \approx c/v$ at a given time.  We call this the `luminosity shell.'

The post-breakout emission proceeds in three phases \citep{nakar10}.  First, during the breakout, the energy of the breakout shell, $\Ebo$, is smeared over the diffusion time of the breakout shell, $\tbo$, producing a roughly constant luminosity of $L \approx \Lbo$, where
\begin{eqnarray}
\label{Lbodef}
\Lbo \approx \dfrac{\Ebo}{\tbo}.
\end{eqnarray}
In reality, the values of $\Lbo$ and $\Ebo/\tbo$ differ by a numerical factor which depends on the fraction of $\Ebo$ which is radiated in one diffusion time, $\tbo$ \citep[see, e.g., ][for a nuanced discussion of this topic]{sapir11,katz12}.  For simplicity, we do not include this factor, as its exact value is not important for our results.

Next, there is a planar evolution which lasts until the breakout shell has doubled in radius, i.e. for $\tbo < t < \ts$.  Here, we assume \citep[following ][]{nakar10} that the radiation originates from the same mass coordinate throughout the planar phase \citep[see, however, ][for discussion of logarithmic corrections which arise when this assumption is dropped]{faran19}. Throughout this phase, the radius of the breakout shell is roughly constant, and because the optical depth does not change appreciably, the breakout shell continues to satisfy $\tau \approx c/\vbo$ (in other words, the breakout shell is also the luminosity shell).  However, the thickness of the breakout shell grows in proportion to $t$.  Accordingly, the breakout shell suffers adiabatic losses with $E \propto t^{-1/3}$ (assuming an adiabatic index of 4/3 , as appropriate for a gas with radiation dominated pressure), and the time to diffuse through the shell increases as $t_d \propto t$.   Thus, the luminosity scales as $L \propto E/t_d \propto t^{-4/3}$.  

\begin{figure}
\centering
\includegraphics[width=\columnwidth]{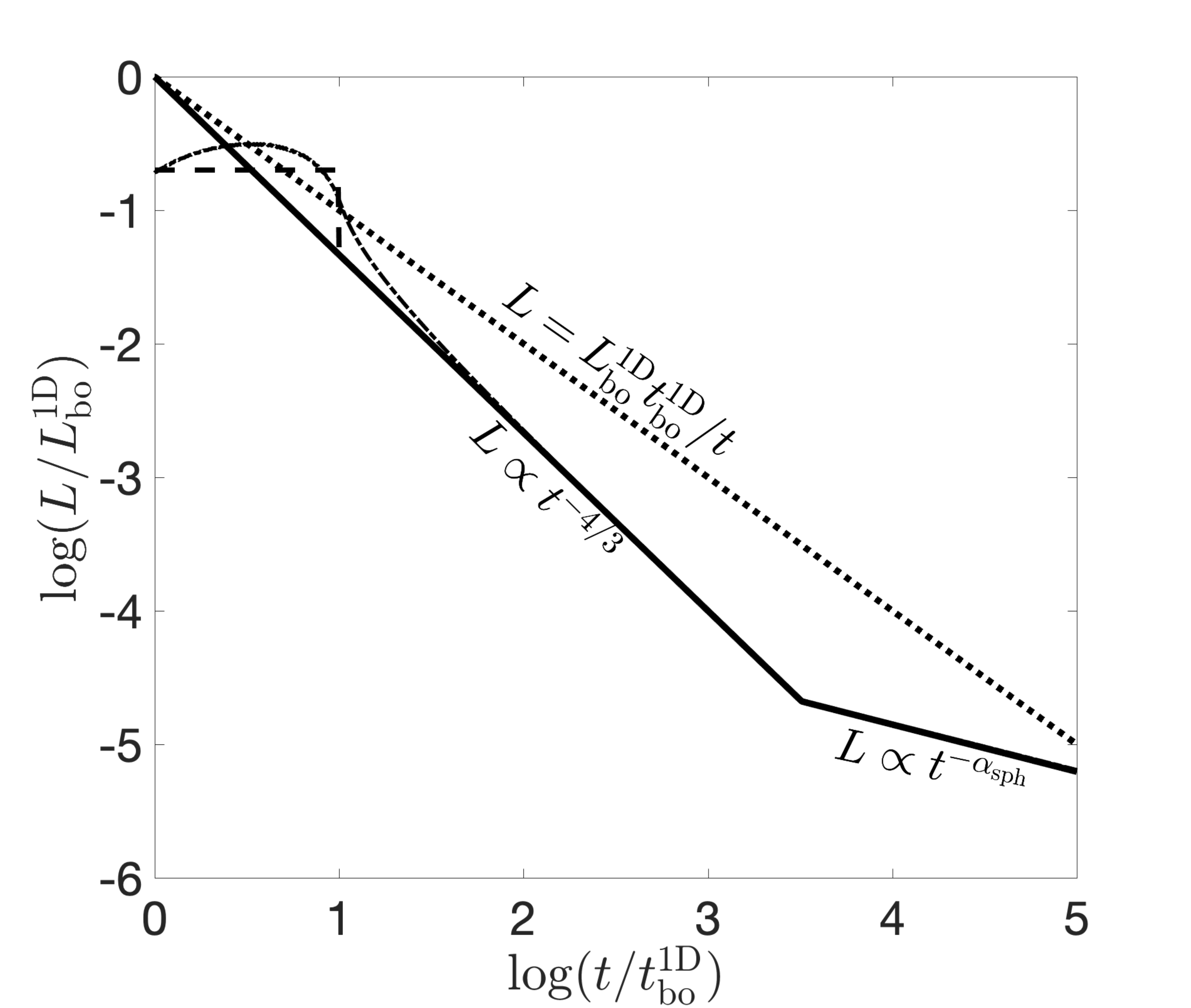}
\caption{Schematic light curve for a spherically symmetric breakout, showing the constant-luminosity plateau phase, planar cooling phase, and spherical cooling phase as indicated by equation~\ref{lightcurvespherical}.  If the light-crossing time $R_*/c$ is longer than the diffusion time of the breakout shell $\tbo$, then the emission is smeared over a light-crossing time, resulting in a longer and fainter breakout as shown by the dashed lined for $R_*/c=10 \tbo$. The dot-dashed line shows a more sophisticated model, which includes light travel time effects, accounts for the reduction in projected area at high latitudes, and sums the contributions from freshly shocked ejecta and ejecta which have already started to cool.  (For further discussion of light-travel time and projected area effects, see Section~\ref{sec:lighttravel}).   Because of the additional contribution from cooling material, the dot-dashed line lies above the dashed line (which neglects this emission).  Regardless of the duration, the total energy released during the breakout flash is $\approx \Lbo \tbo$.  A dotted line of constant radiated energy, $Lt=\Lbo\tbo$, is shown for reference.}
\label{fig:Lightcurve1D}
\end{figure}

Finally, for $t > \ts$ there is a spherical phase during which the adiabatically cooling material interior to the breakout shell is gradually revealed.  Because the distribution of energy in the ejecta is governed by shock acceleration near the edge of the star, the evolution in this case depends on the power-law index $n$ describing the star's outer density profile.  The overall light curve found by connecting these three phases is  \citep{nakar10}
\begin{eqnarray}
\label{lightcurvespherical}
L_{\rm sph}(t) \approx 
\begin{cases}
\Lbo, & t<\tbo \\
\Lbo \left(\dfrac{t}{\tbo}\right)^{-4/3}, & \tbo < t < \ts \\
\Lbo \left(\dfrac{\ts}{\tbo}\right)^{-4/3} \left(\dfrac{t}{\ts}\right)^{-\alpha_{\rm sph}}, & t > \ts
\end{cases}
\end{eqnarray}
where
\begin{eqnarray}
\label{alphas}
\alpha_{\rm sph} = \dfrac{2(6\mu n-1)}{3(1+n+\mu n)} \approx 0.35,
\end{eqnarray}
as shown in Figure~\ref{fig:Lightcurve1D}. The numerical value of $\alpha_{\rm sph}$ was obtained for $\mu=0.19$ and $n=3$.

In order to set the stage for our discussion of light travel time in non-spherical explosions (which will be considered in Section~\ref{sec:lighttravel}), we also briefly review here the behaviour in the spherical case when the light--crossing time of the star is long compared to the diffusion time of the breakout layer. In this $R_*/c > \tbo$ limit, the initial breakout flash is smeared over a time-scale of $R_*/c$.  The emission from angle $\theta$ does not reach the observer until the time $t=(R_*/c)(1-\cos\theta)$.  For $\theta \ll 1$, $t\approx R_*\theta^2/2c \propto \theta^2$.  Furthermore, as the energy released from a patch of angular size $\theta$ scales as $E \approx \Ebo \theta^2 \propto \theta^2$ in the small-angle limit, the luminosity $L\propto E/t$ is approximately constant at early times, with an initial value given by $L \approx E/t \approx 2\Lbo \tbo(R_*/c)^{-1}.$

A simple approximation for the light curve in this case is to treat $L$ as constant all the way up to $R_*/c$, as shown by the dashed line in Figure~\ref{fig:Lightcurve1D}. The light curve may be refined even further by considering two additional effects.  First, regions of the stellar surface which are far from the line of sight contribute significantly less to the total observed emission, due to having a smaller projected area (see Section~\ref{sec:lighttravel} for more discussion of projected area).  This geometrical effect reduces the luminosity at $t\sim R_*/c$ and smooths out the transition between the $L\propto t^0$ and $L\propto t^{-4/3}$ segments of the light curve. 
Secondly, for $\tbo<R_*/c$, the cooling emission produced by material along the line of sight starts to reach the observer well before higher latitudes become visible.  At a given time, the total luminosity is the sum of the breakout emission from $\theta\approx (2ct/R_*)^{1/2}$, and the cooling emission originating from smaller $\theta$ (for more about the relative contribution from freshly-heated and already-cooling material, see Section~\ref{sec:quasispherical}).    The light curve when these two additional effects are accounted for can be calculated using, e.g., equation 30 in \citet{katz12}, with $L(t)$ given by our equation~\ref{lightcurvespherical}. The result is shown by the dot-dashed line in Figure~\ref{fig:Lightcurve1D}.  As expected, it converges to the value given in equation~\ref{lightcurvespherical} when $t\gg R_*/c$.  At a few $\tbo$, the light curve exceeds the simple estimate by a factor of $\sim 2$ due to the added contribution from the cooling material.

Note that at times $t<\tbo$, the light curve exhibits an exponentially rising behaviour, as photons diffusing far ahead of the shock start to escape.  At a (negative) time $t$ just before the breakout, when the shock is a distance $x\approx-\vbo t \propto t$ from the stellar surface, a fraction $e^{-x^2}$ of the photons are able to diffuse ahead of the shock by a distance $x$ and escape, leading to an exponentially rising light curve $L \propto e^{-t^2}$ at the earliest times \citep[e.g.,][]{sapir11,shussman16}.  As we do not treat the rising phase of the light curve, our results are only valid for $t\ga \tbo$.

\subsection{Non-instantaneous quasi-spherical plane-parallel breakout}
\label{sec:quasispherical}
Since the breakout emission is smeared over the photon diffusion time, the resulting light curve is indistinguishable from a purely spherical breakout as long as the time for the breakout to cross the star, $\twrap$, is shorter than $t_{\rm bo}^{\rm 1D}$. If however $\twrap > t_{\rm bo}^{\rm 1D}$, the breakout is effectively instantaneous up to an angle $\theta_{\rm bo}$, reached at $t_{\rm bo}^{\rm 1D}$, while at later times the shock continues to sweep the rest of the stellar surface, and the breakout's non-simultaneity becomes apparent, with the breakout duration set by $\twrap$. 

Approximating the breakout as a sequence of radial breakouts from annuli of increasing latitude, the full light curve can be obtained by integrating the emission across the stellar surface. The breakout reaches an angle $\theta$ at time $t_0(\theta)$, as given in equation \ref{ttheta}. The annulus covering the interval $(\theta,\theta+d\theta)$ subtends a solid angle $2\upi \, \sin{\theta} \, {\rm d}\theta$, and has a projected area of $2\upi R_*^2 \sin \theta \cos \theta {\rm d}\theta$ (see Section~\ref{sec:lighttravel} for further discussion).  Comparing this with the total projected area of the star $\upi R_*^2$, we find that, at time $t$, the annulus contributes an emission of $2 L_{\rm sph}(t-t_0(\theta)) \sin \theta \cos \theta {\rm d}\theta = L_{\rm sph}(t-t_0(\theta)) \sin(2\theta){\rm d}\theta$, where $L_{\rm sph}$ is given by equation~\ref{lightcurvespherical}, and $L_{\rm sph}(t<0)\equiv0$. Integrating the contributions from all annuli, we have
\begin{equation} \label{eq:L_conv_t_planar_non_inst}
    L(t) = \int_0^{\upi/2} { L_{\rm sph}(t-t_0(\theta)) \, \sin{2\theta} \, d\theta \,. }
\end{equation} 
As expected, the luminosity satisfies $L(t)\rightarrow L_{\rm sph}(t)$ as $t \rightarrow \infty$.

In the following paragraphs we make a few simplifying approximations to obtain the light curve as a broken power-law instead of the above integral expression. First, we concentrate on small latitudes, $\theta \ll 1$, where $\sin{\theta} \approx \theta$ and $\cos{\theta} \approx 1$. Additionally, instead of integrating across the entire stellar surface, we consider only the latitudes that dominate the emission at any given time $t$. As we later show, at times $\tbo < t < \twrap$, the emission is dominated by the material that is currently breaking out, i.e., the material at latitude $\theta$ satisfying $t_0(\theta) \approx t$.

The luminosity of the initial phase in this regime is set by the amount of energy enclosed in the breakout layer up to $\theta_{\rm bo}$, emitted over the diffusion time $t_{\rm bo}^{\rm 1D}$. Assuming $\theta_{\rm bo} \ll 1$, the projected area of this patch is a fraction $(\upi R_*^2 \theta_{\rm bo}^2)/(\upi R_*^2) = \theta_{\rm bo}^2$ of the star's total projected area.  As a result, the luminosity at this stage is given by
\begin{equation}
    L(t<t_{\rm bo}^{\rm 1D}) = L_{\rm bo}^{\rm 1D} \theta_{\rm bo}^2 \,.
\end{equation}
At later times, $t>t_{\rm bo}^{\rm 1D}$, the energy contained in the breakout layer up to angle $\theta(t)$ is proportional to $\theta^2$, corresponding to the area of a spherical cap of opening angle $\theta \ll 1$, and the emission duration is $t$. The luminosity therefore scales as $L\propto \theta^2 / t \propto t^{(1-k)/(1+k)}$. Note that as before, the projected area is damped by a factor $\cos{\theta}$. We currently neglect this factor, and defer this discussion to section \ref{sec:lighttravel}.

Every shock heated surface element begins to expand and cool adiabatically after time $t_{\rm bo}^{\rm 1D}$ has elapsed since it was first shocked. Since the ejected material expands radially, the volume of every element increases linearly with the time, in times shorter than $\ts$. Assuming $\gamma=4/3$, the internal energy of each element decreases as $t^{-1/3}$, and the luminosity of the adiabatically cooling material scales as $t^{(1-k)/(1+k)-1/3}$. However, since the breakout continues up to time $\twrap$, newly-heated surface elements also contribute to the cumulative light curve. Since $(1-k)/(1+k) > (1-k)/(1+k)-1/3$ is always satisfied, the contribution of the freshly shocked material currently breaking out always outshines the adiabatic cooling emission from previously ejected material.  

After the shock has wrapped around the stellar surface, the most recently ejected material cools adiabatically on a time-scale of ${t_{\rm bo}^{\rm 1D} \ll \twrap}$, and the luminosity sharply drops to match the light curve of the adiabatically-cooling planar phase of the effectively 1D case.  Put together, the luminosity in this phase is given by the following expression, which is a simplified approximation of the complete integral of equation \ref{eq:L_conv_t_planar_non_inst}:
\begin{eqnarray}
\label{lightcurves_planarnoninst}
L(t) \approx 
\begin{cases}
\Lbo \theta_{\rm bo}^2, & t<\tbo \\
\Lbo \theta_{\rm bo}^2 \left(\dfrac{t}{\tbo}\right)^{(1-k)/(1+k)}, & \tbo < t < \twrap \\
L_{\rm sph}(t) , & t > \twrap
\end{cases}.
\end{eqnarray}
In the spherical limit, where $\epsilon \to 0$ or $k \to \infty$, $\theta_{\rm bo} \to \upi/2$, and $\twrap \to 0$. The intermediate interval $t_{\rm bo}^{\rm 1D} < t < \twrap$ does not exist, and we obtain the same light curve as in the spherical case.

\begin{figure}  
\centering
\includegraphics[width=\columnwidth]{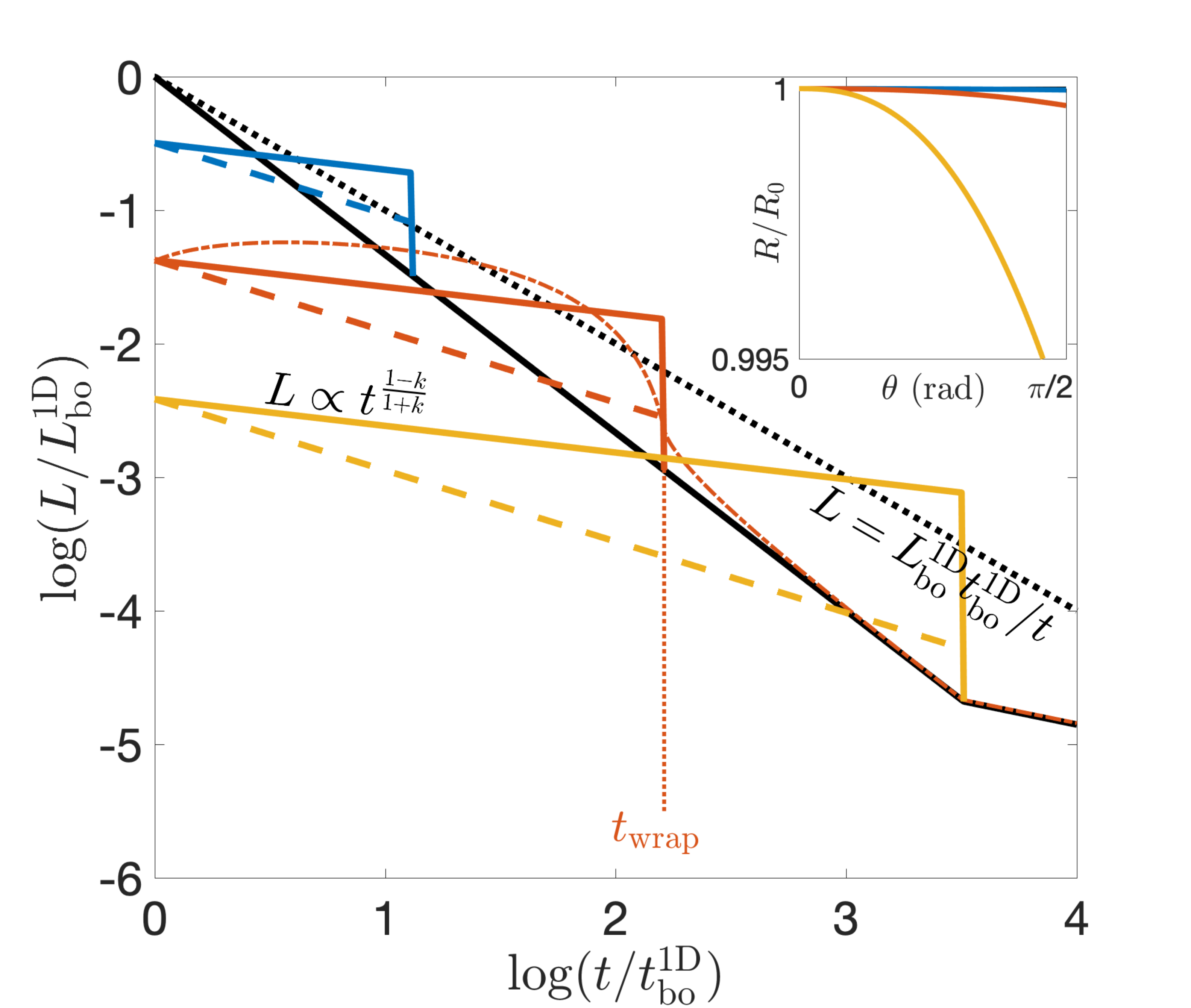}
\caption{Schematic light curve for a non-instantaneous plane-parallel breakout (coloured lines), compared to the fully spherical case (black line), adopting $\vstar/\vbo=0.01$, 
and $\epsilon=4\times10^{-5}$ (blue), $5\times10^{-4}$ (red), or $0.01$ (yellow).  The  duration of the plateau phase is the time for the breakout point to wrap around the star, $\twrap$ (labelled  with a dotted vertical line for the red curve).  Along the black dotted line, the radiated energy $\approx \Lbo \tbo$ is constant.  The dashed lines show the expected emission from adiabatically-cooling material that has already broken out at times $t < \twrap$; in all cases this emission is outshone by emission from the currently-breaking-out material.  After the breakout ends, the ejecta rapidly cool adiabatically on a time-scale of $\sim \tbo \ll \twrap$, and from then on the emission is the same as in the adiabatically-cooling phase of the spherical light curve.  The thin dot-dashed red line shows, for the $\epsilon=5\times 10^{-4}$ case, the light curve obtained from the full integral expression in equation~\ref{eq:L_conv_t_planar_non_inst}.   \textit{Inset:} The shock's initial radius (in units of $R_0$) as a function of $\theta$ for each model, at the time when shock acceleration begins.  We point out that even very small deviations from a spherical shock (note the scale of the $y$-axis) result in appreciable differences in the light curve.}
\label{fig:Lightcurve_NonInstPlaneParallel}
\end{figure}

Figure \ref{fig:Lightcurve_NonInstPlaneParallel} summarizes the light curves obtained in this regime. Note that the value of $\vstar/\vbo=0.01$ used in the figure is chosen only for clarity of presentation, to increase the separation between the curves.  Under realistic conditions, $\vstar/\vbo=0.1$ is a more typical value.  As expected, the breakout releases approximately the same total energy as in the spherical case, but over a longer time-scale $\twrap$.  At times $t>\tbo$, the total luminosity decays more slowly than the cooling emission of previously ejected material (solid versus dashed lines in the figure). The figure also compares the accurate light curve calculated via the integral in equation~\ref{eq:L_conv_t_planar_non_inst} to the crude analytical estimate given by equation~\ref{lightcurves_planarnoninst} for the $\epsilon=5\times 10^{-4}$ case.  The integrated light curve (dot-dashed red curve) is brighter than the simple approximation by a factor of 2--3 in the interval ${\tbo < t < \twrap}$, as it includes contributions both from the material currently breaking out, and from material which broke out earlier and is now cooling.  In addition, the sharp transition at $t=\twrap$ which appears in the analytical model is smoothed out by geometrical effects.  None the less, the rough approximation still captures the time-scale and luminosity of the initial breakout pulse reasonably well.  At late times, as expected, the integrated light curve asymptotically approaches the light curve of a spherically symmetric explosion (solid black line).

We note that the simple approximation in equation~\ref{lightcurves_planarnoninst} overestimates the total energy released during the initial breakout flash, because it does not account for the reduced projected area at high latitudes.  Using equations~\ref{twrap} and~\ref{thetabo} to replace $\theta_{\rm bo}$ and $\twrap$ in equation~\ref{lightcurves_planarnoninst}, we find that the total emitted energy is $ L(\twrap)\twrap \approx (\upi/2)^2 \Lbo \tbo$, a factor of $(\upi/2)^2$ larger than the true value of $\Lbo \tbo$.  The reason for the factor of $(\upi/2)^2$ is that the simple estimate effectively assumes that the small-angle approximation applies all the way up to $\theta=\upi/2$ and $t=\twrap$.  Under this assumption, the projected area at $\theta=\upi/2$ is $\upi R_*^2 \theta^2 = \upi(\upi/2)^2 R_*^2$, which is larger than the actual projected area $\upi R_*^2$ by a factor of $(\upi/2)^2$. As a result, at $t=\twrap$ the solid coloured curves lie above the dotted $L=\Lbo(\tbo/t)$ line by a factor of $(\upi/2)^2= 2.46$. Despite the deficiencies of the simplified model at $t\approx \twrap$, it has the advantage that it predicts the correct luminosity at $t=\tbo$, and agrees with the complete integral expression to within a factor of $\sim 2$ for $t\la 0.7 \twrap$.

\subsection{Non-instantaneous quasi-spherical oblique breakout}
\label{sec:quasioblique}

As the explosion's asphericity gradually increases, the breakout duration increases, as described in the previous section. However, when $\twrap > \ts$, the breakout's pattern speed drops below $\vbo$ somewhere along the stellar surface, marking the transition from a non-instantaneous, yet locally plane-parallel breakout, to an oblique breakout. When the breakout is oblique, the maximal ejecta velocity is limited to the pattern speed (rather than the 1D breakout velocity), as discussed in Section~\ref{sec:velocityscales}. Additionally, in this regime, the breakout results in a significant non-radial ejecta flow that wraps around the stellar surface.

As discussed in Section~\ref{sec:breakoutclasses} and shown in Figure~\ref{fig:SchematicBreakoutClasses}, once the breakout becomes oblique, the breakout location is blocked from view. Therefore, only the material within the plane-parallel breakout region (i.e., within a spherical cap extending to an angle of $\theta_{\rm obq}$ from the axis) contributes to the breakout's prompt emission. Consequently, the energy radiated during the initial breakout flash is only $\approx \Ebo \theta_{\rm obq}^2$.  The duration of the flash is also shortened: instead of being the time for the shock to cross the star, it is given by the time for the shock to cross the parallel breakout region, which is
\begin{equation} \label{ttr}
    t_{\rm obq} \approx R_*\theta_{\rm obq}/v_{\rm p}(\theta_{\rm obq})=R_* \theta_{\rm obq}/\vbo = \ts \theta_{\rm obq}\, .
\end{equation}

\begin{figure}
\centering
\includegraphics[width=\columnwidth]{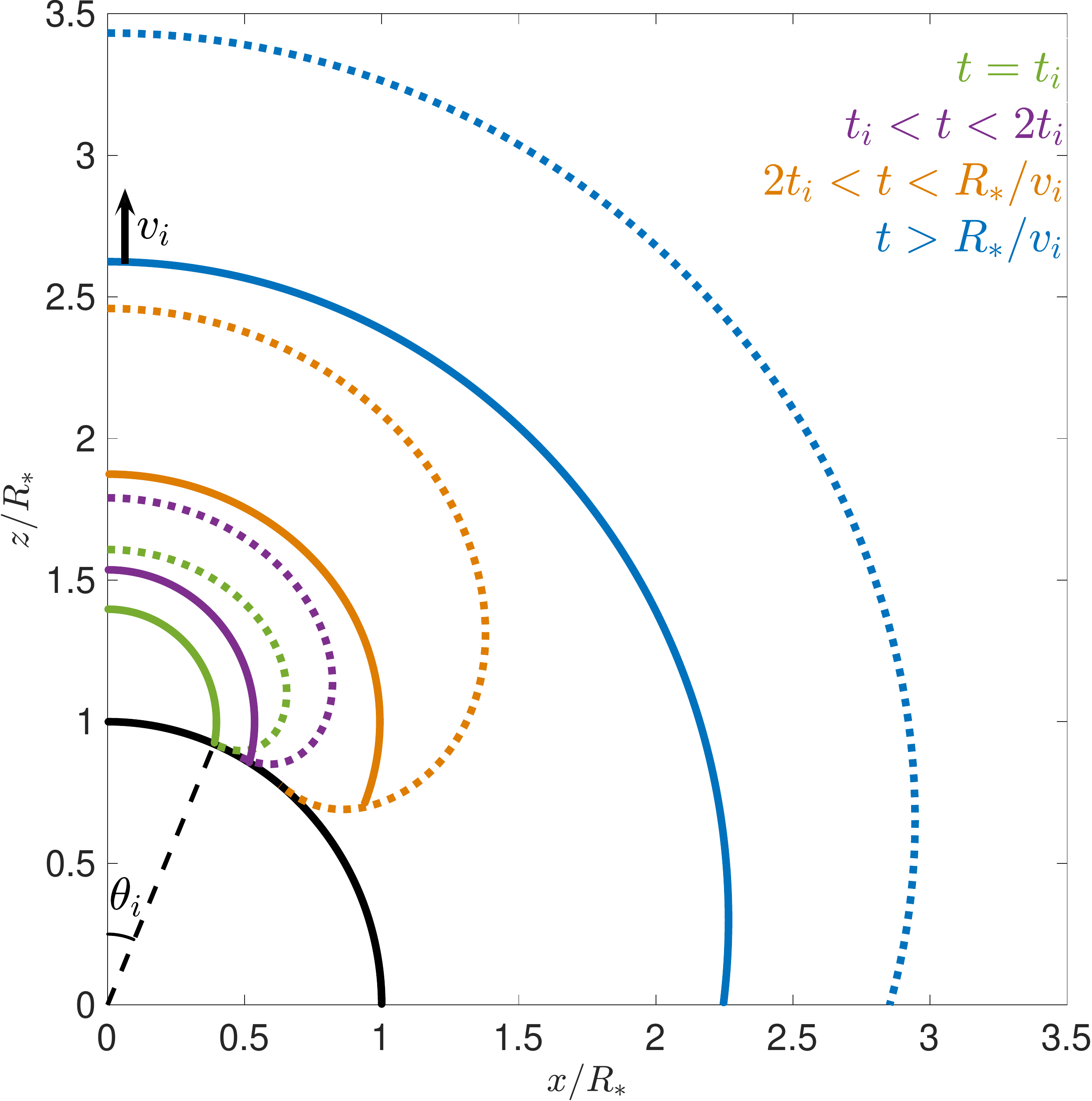}
\caption{Dynamical evolution of a shell ejected during an oblique breakout. The black curve is the surface of the star.  The other colours show the shape of the ejecta at four different times.  At each time, the outer edge of the ejecta cloud is shown with a dotted line, while a shell of constant velocity $v=v_i$ is drawn with a solid line.  \textit{Green:} The initial conditions of the shell with velocity $v_i$ when it is ejected at $t=t_i$.  The initial angular size of the shell, $\theta_i$, is indicated by the black dotted line.  \textit{Purple:} The early planar evolution of the shell for times $t<2t_i$.  The angular size of the shell is still approximately $\theta_i$, and it extends to a radius of $R_* +\theta_i R_* \approx R_*$ from the origin.  \textit{Orange:} The sideways expanding phase of shell evolution for $2t_i < t < R_*/v_i$.  The angular extent of the shell is now significantly larger than $\theta_i$.  However, the shell is still contained within a radius of $\la 2R_*$.  \textit{Blue:} Spherical phase of shell evolution for $t>R_*/v_i$.  The shell now subtends an angle of $\sim 1$ radian, and its radius is substantially larger than $R_*$.}
\label{fig:ShellEvolution}
\end{figure}

After this initial flash, the observed signal is determined not by the emission from the breakout point, but rather by the emission from the adiabatically cooling, sideways-expanding, mushroom-shaped ejecta cloud.  We estimate the emission from this cloud in the following way.  Since the velocity of a given ejecta element remains roughly constant after breakout {(it changes roughly by a factor of of two or less)}, we first divide the ejecta into discrete constant-velocity shells, each labelled  by an index $i$ and containing material with initial velocities between $v_i$ and $v_i/2$ (here and below, the subscript $i$ is used to indicate the initial properties of the $i$-th shell). Figure~\ref{fig:ShellEvolution} shows the evolution of one such shell. Since the maximum ejecta velocity is limited to the pattern speed in an oblique breakout, at a given location on the surface, the speed of the fastest material ejected by the breakout is given by $v_{\rm p}(\theta)$. The initial opening angle of the $i$-th shell, $\theta_i$, therefore satisfies $v_{\rm p}(\theta_i)=v_i$.  The time it takes the shock to reach $\theta_i$, crossing a distance $R_* \theta_i$, is $t_i\approx R_*\theta_i/v_i$.  In the same time, the on-axis material with the same speed $v_i$ travels the same distance, $v_i t_i \approx R_*\theta$.  Therefore, we can approximate the initial shape of each shell as a spherical arc with radius $\approx R_* \theta_i$, centred at the point $(r=R_*$,$\theta=0)$, as depicted the green solid curve in Figure~\ref{fig:ShellEvolution}.

Next, we consider the dynamical evolution of such a shell, ejected at time $t_i$ and expanding at constant velocity $v_i$.  Initially, there is a planar phase (shown in purple in Figure~\ref{fig:ShellEvolution}) lasting until $\approx 2 t_i$, during which the shell's radius (as measured from the origin) and angular extent do not vary significantly, i.e. $r\approx R_*$ and $\theta \approx \theta_i$.  After $t \gtrsim 2t_i$, sideways expansion becomes significant (see the orange curve in Figure~\ref{fig:ShellEvolution}), and its angular size increases as $\theta(t) \sim v_i t/R_* \propto t$; however, each point along the shell still has a radius of $r \approx R_*$ during this phase.  Finally, once $t>R_*/v_i=t_i/\theta_i$, the shell's angular extent becomes $\sim 1$ radian, and its radius is of order $\sim 2 R_*$.  Beyond this time, the shell enters a spherical expansion phase, where $\theta$ is constant while $r \approx v_i t \propto t$, as depicted by the blue curve in Figure~\ref{fig:ShellEvolution}.  

Finally, once the dynamics of each shell are known, we estimate the time-dependent luminosity produced by the ensemble of shells.  At any given time, we make the approximation that the luminosity is dominated by the shell whose optical depth satisfies $\tau(t)=c/v_i$, such that its radiation has just become able to diffuse out.  Following \citet{nakar10}, we call this shell the `luminosity shell,' and notate variables describing its properties using a `$\sim$' symbol.  (Note that, for this rough estimate, we assume purely radial diffusion. If diffusion in the non-radial direction is significant, the emission might be altered somewhat from our results, particularly for off-axis observers.  However, including the detailed effects of non-radial diffusion is beyond the scope of this work.)  To calculate the optical depth of a given shell, we first notice that since the shell's mass is constant, its optical depth $\tau \approx \kappa m_i/(4\upi r^2 \theta^2)$ is inversely proportional to its surface area $(r\theta)^2$.  Then, considering each dynamical regime discussed above, we obtain that the optical depth is constant in the planar phase, and declines as $\tau\propto t^{-2}$ in both the sideways-expanding and the spherical expansion phases.  A general expression for the optical depth of the $i$-th shell at time $t$ is then given by
\begin{equation}
\label{tauevolution}
    \tau(\theta_i,t) = \tau_i(\theta_i) \left( \frac{t}{t_i(\theta)} \right)^{-2} \,,
\end{equation}
where the initial optical depth of the shell $\tau_i$ is to be determined below.

To complete the derivation of the light curve, we must first work out the initial properties of each shell. We choose to express these properties for each shell as a function of that shell's initial opening angle, $\theta_i$.  Note that the fact that each shell has a distinct initial angular size is a key difference from the spherically symmetric case (where all shells subtend the full $4\upi$ steradians); this affects how the shell's mass and energy scale with its velocity, and thus alters the light curve.  Once the initial properties of all the shells are known, we then locate the luminosity shell at a given time. Finally, by determining the luminosity shell's properties, we estimate how much energy is available in the luminosity shell to be radiated, and compute the luminosity at time $t$.  

In what follows, to keep the exponents appearing in equations relatively simple, we adopt nominal values of $\mu=0.19$ for the shock acceleration parameter, and $n=3$ for the power-law index of the star's outer density profile, and retain only the dependence on the parameter $k$ characterizing the angular dependence of the shock asphericity.  For completeness, the full expressions for arbitrary $\mu$ and $n$ are also given in Appendix~\ref{appendixb}.

To begin with, we recall that after the breakout becomes oblique, at $\theta > \theta_{\rm obq}$, the ejecta velocity is limited by the breakout's pattern speed, i.e. $v_i=v_{\rm p}(\theta_i)$.  The time $t_i$, at which the shock reaches $\theta_i$ and the shell with velocity $v_i$ is ejected, is then found using equation~\ref{vp}.  This yields
\begin{equation}
    \label{ti}
    t_i(\theta_i) \approx \frac{R_* \theta_i}{v_{\rm p}(\theta_i)} = t_{\rm obq} \left( \frac{\theta_i}{\theta_{\rm obq}} \right)^{1+k} \,,
\end{equation}
where $t_{\rm obq}$ is the time at which the breakout first becomes oblique (given by equation~\ref{ttr}), and the non-instantaneous plane-parallel phase terminates.  In this regime $\tbo < t_{\rm obq} < \ts$.  The corresponding velocity at this time is
\begin{equation} \label{eq:v_theta}
    v_i(\theta_i) = v_{\rm p}(\theta_i) = \vbo \left( \frac{\theta_i}{\theta_{\rm obq}} \right)^{-k} \,,
\end{equation}
where we used the fact that $v_{\rm p}(\theta_{\rm obq})=\vbo$.

The mass of the shell whose initial angular extent is $\theta_i$ can be approximated by the mass of all elements with velocity larger than $v_i(\theta_i)$. 
Following the solution of \citet{sakurai60}, density scales as a power of the shock velocity $\rho \propto v^{-1/\mu}$. Using equation \ref{eq:v_theta} and scaling by the breakout density dictated by the 1D theory, we find the initial density of a mass shell with an initial angle $\theta_i$ is
\begin{equation}
    \rho_i(\theta_i) = \rho_{\rm bo}^{\rm 1D} \left( \frac{v_i(\theta_i)}{\vbo} \right)^{-1/\mu} = \rho_{\rm bo}^{\rm 1D} \left( \frac{\theta_i}{\theta_{\rm obq}} \right)^{5.26k} \,,
\end{equation}
while its corresponding width, $y_i$, is related to the density through
\begin{equation} \label{eq:yi_theta}
    y_i(\theta_i) = y_{\rm bo}^{\rm 1D} \left( \frac{\rho_i(\theta_i)}{\rho_{\rm bo}^{\rm 1D}} \right)^{1/n} = y_{\rm bo}^{\rm 1D} \left( \frac{\theta_i}{\theta_{\rm obq}} \right)^{1.75k} \,.
\end{equation}
The mass enclosed in a shell is given by the density $\rho(\theta)$ multiplied by the shell's volume,
\begin{equation} \label{eq:m_theta}
    m_i(\theta_i) = \rho_i(\theta_i) (R_* \theta_i)^2 y_i(\theta_i) = m_{\rm obq} \left( \frac{\theta_i}{\theta_{\rm obq}} \right)^{7.02k+2} \,,
\end{equation}
where $m_{\rm obq} = R_*^2 \theta_{\rm obq}^2 \rho_{\rm bo}^{\rm 1D} y_{\rm bo}^{\rm 1D}$ is the breakout mass enclosed in the region where the breakout is plane-parallel.

The 1D breakout layer's optical depth is, by definition $c/\vbo$. The initial optical depth of the $i$-th shell is thus given by
\begin{equation}
    \label{taui}
    \tau_i(\theta_i) \approx \kappa \rho(\theta_i) y_i(\theta_i) \approx \frac{c}{\vbo} \left( \frac{\theta_i}{\theta_{\rm obq}} \right)^{7.02k} \,.
\end{equation}

Finally, the shell's initial internal energy at the time of breakout, before it has cooled adiabatically, is similar to its kinetic energy,
\begin{equation} \label{eq:Ei_theta}
    E_i(\theta_i) \approx m_i v_i^2 \approx E_{\rm obq} \left( \frac{\theta_i}{\theta_{\rm obq}} \right)^{5.02k+2} \,,
\end{equation}
where $E_{\rm obq} \approx m_{\rm obq} (\vbo)^2 \approx \Ebo \theta_{\rm obq}^2$ is the energy contained in the region in which the breakout is plane-parallel.

A given shell becomes visible when its optical depth satisfies $\tau(\theta_i,t) = c/v_i(\theta_i)$. Solving for the luminosity shell's initial angle $\Tilde{\theta}_i$ at time $t$ by using equations~\ref{tauevolution}, \ref{ti}, and~\ref{taui}, we obtain
\begin{equation} \label{eq:aspherical_theta_t}
    \Tilde{\theta}_i = \theta_{\rm obq} \left( \frac{t}{t_{\rm obq}} \right)^{0.25/(k+0.25)} \,.
\end{equation}
The visible shell's velocity, mass, initial width and initial internal energy are obtained by plugging in $\theta_i=\tilde{\theta}_i(t)$ in equations \ref{eq:v_theta}, \ref{eq:m_theta}, \ref{eq:yi_theta} and \ref{eq:Ei_theta}, and denoting the properties of the currently visible shell as $\tilde{v}_i$, $\tilde{m}_i$, $\tilde{y}_i$ and $\tilde{E}_i$, respectively.

We now take into account the adiabatic losses the luminosity shell undergoes as it expands, before becoming visible. Its volume increases from $\Tilde{V}_i \approx R_*^2 \Tilde{\theta}_i^2 \Tilde{y}_i$ to roughly $\Tilde{V}_f \approx (\Tilde{v}_i t)^3$ at late times ($t \gg \Tilde{y}_i/\Tilde{v}_i$). For an adiabatic index $\gamma$, the internal energy of the luminosity shell at time $t$ is given by
\begin{equation}
    \Tilde{E}_f = \Tilde{E}_i \left( \frac{\Tilde{V}_f}{\Tilde{V}_i} \right)^{1-\gamma} = E_{\rm obq} \left( \frac{t_{\rm bo}}{t_{\rm obq}} \right)^{1/3} \, \left( \frac{t}{t_{\rm obq}} \right)^{\frac{0.65k+0.42}{k + 0.25}} \,,
\end{equation}
where in the second equality we used $t_{\rm obq} = R_* \theta_{\rm obq}/\vbo$,  $t_{\rm bo}=y_{\rm bo}^{\rm 1D}/v_{\rm bo}$, and $\gamma=4/3$ as appropriate for a radiation-dominated gas. 

The bolometric luminosity is therefore given by
\begin{equation} \label{eq:L_asph_oblique}
    L_{\rm asph}(t) \approx \Tilde{E}_f / t = \frac{E_{\rm obq}}{t_{\rm obq}} \left( \frac{t_{\rm bo}}{t_{\rm obq}} \right)^{1/3}
    \, \left( \frac{t}{t_{\rm obq}} \right)^{-\alpha_{\rm asph}(k)} \,,
\end{equation}
where 
\begin{eqnarray}
    \alpha_{\rm asph}(k) & = & \frac{2[k(6\mu n - 1) - 2 \mu n]}{3[k(1+n+\mu n) + 2\mu n]} \\
    \nonumber & \approx & \frac{0.35 k - 0.17}{k + 0.25} \,,
\end{eqnarray}
and again we used $n=3$ and $\mu=0.19$.

The exponent $-\alpha_{\rm asph}(k)$ can be either positive or negative. The critical value of $k$ separating rising and falling light curves, for which $L_{\rm asph}=\rm const.$ is $k_{\rm crit} = 2\mu n / (6\mu n - 1 ) \approx 0.47$.
As expected, in the limit $k \to \infty$, $\alpha_{\rm asph}(k) \to \alpha_{\rm sph}$, reproducing the same late-time behaviour seen in the spherically symmetric theory.

How does the light curve of an aspherical explosion considered here compare with the luminosity of a spherically symmetric explosion of the same energy and stellar parameters? Since $E_{\rm obq} = \Ebo \theta_{\rm obq}^2$, and $t_{\rm obq} = \ts \theta_{\rm obq}$, we rewrite equation \ref{eq:L_asph_oblique} in terms of the luminosity of a similar spherical explosion
\begin{equation}
\label{lightcurves_obliquenoninst}
    L_{\rm asph}(t) = L_{\rm sph}(t) \left( \frac{t}{t_s} \right)^{\alpha_{\rm sph} - \alpha_{\rm asph}(k)} \theta_{\rm obq}^{\alpha_{\rm asph} + 2/3} \,,
\end{equation}
where $L_{\rm sph}(t)$ is the luminosity given in equation \ref{lightcurvespherical} at late times $t>t_s$. An aspherical explosion is always dimmer than an equivalent spherical explosion at $t=t_s$, but since $\alpha_{\rm sph} \geq \alpha_{\rm asph}(k)$, at time
\begin{equation}
    t_{\rm eq} = \ts \theta_{\rm obq}^{-(\alpha_{\rm asph}(k)+2/3)/(\alpha_{s}-\alpha_{\rm asph}(k))} \,,
\end{equation}
the luminosities satisfy $L_{\rm asph}(t_{\rm eq}) = L_{\rm sph}(t_{\rm eq})$. 

The time $t_{\rm eq}$ can also be interpreted as the time at which the visible shell's initial angle is of order unity, i.e., $\Tilde{\theta}_i(t) \approx 1$ (see equation \ref{eq:aspherical_theta_t}). At this angle the breakout's pattern speed is slowest, and at times $t>t_{\rm eq}$, the emission is dominated by deeper and slower material ($v<v_{\rm p}(\theta\approx 1)$), whose hydrodynamic properties are not influenced by the explosion's asphericity. Once this occurs, the expression ${L=L_{\rm asph}(t)}$ becomes invalid, and we have ${L=L_{\rm sph}(t)}$ instead. 

\begin{figure}
\centering
\includegraphics[width=\columnwidth]{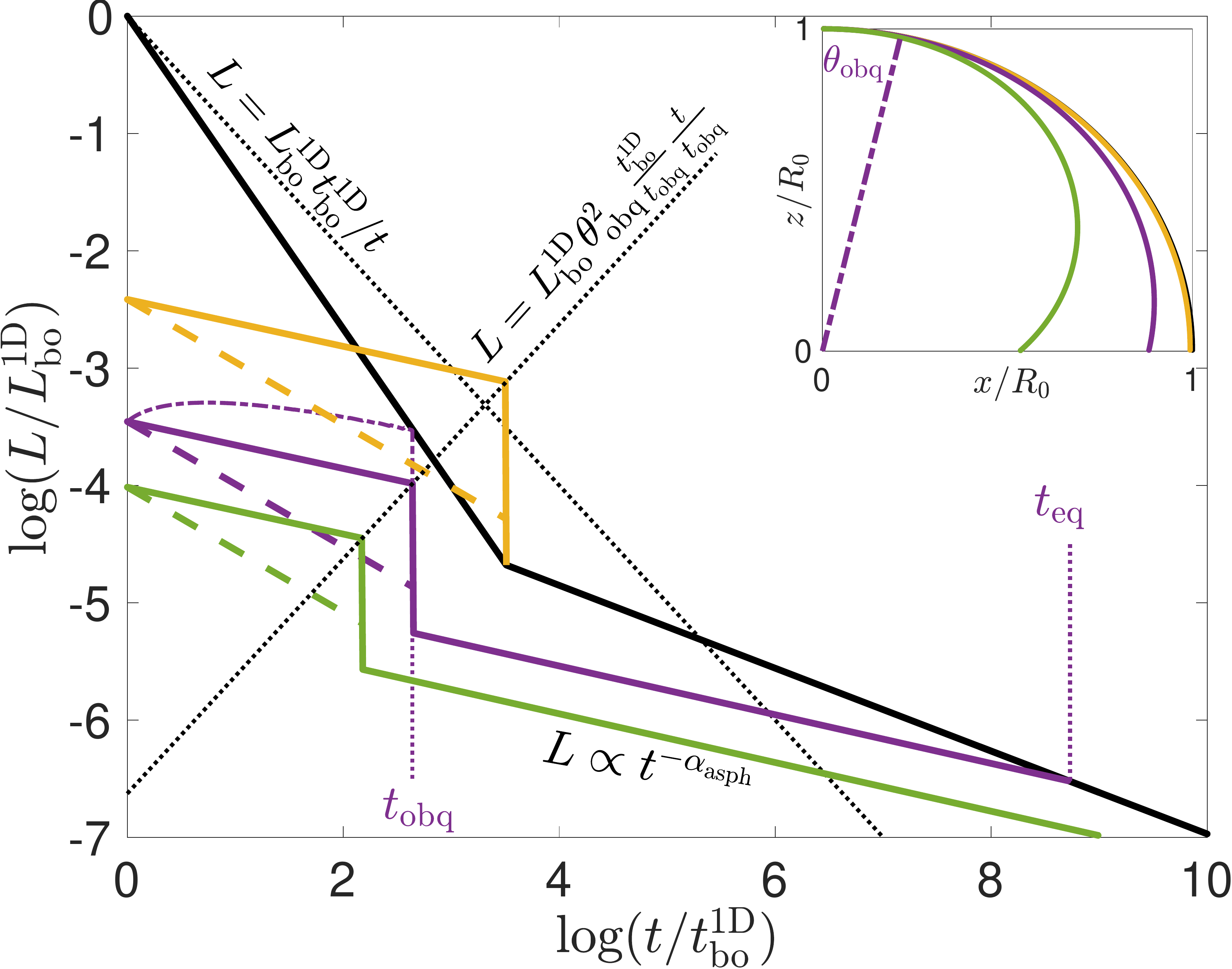}
\caption{Schematic light curve for a non-instantaneous oblique breakout (coloured lines), compared to the fully spherical case (black line), adopting $\vstar/\vbo=0.01$, $k=1.5$, and $\epsilon=0.01$ (yellow), $0.2$ (purple), or $1$ (green).  In this case, the  duration of the plateau phase is the time $t_{\rm obq}$ for the breakout to cross the plane-parallel breakout region of angular size $\theta_{\rm obq}$.  After this time, the rest of the breakout is hidden by the sideways-spreading ejecta cloud. At the end of the plateau phase, the luminosity lies along the upwards-sloping dotted line, along which the observed energy of the breakout flash is $L(t_{\rm obq})t_{\rm obq}=\Lbo \tbo \theta_{\rm obq}^2$. As before, the dashed lines show the expected emission from adiabatically-cooling material that has already broken out during the parallel phase.  Once the breakout flash subsides, the emission is dominated by cooling emission from the sideways-spreading ejecta cloud, which scales as $L\propto t^{-\alpha_{\rm asph}}$.  As time goes by, the observer sees progressively deeper layers of this cloud, until eventually the spherically-expanding ejecta are seen at time $t_{\rm eq}$.  At this point the emission again rejoins the spherically-symmetric solution.  The times $t_{\rm obq}$ and $t_{\rm eq}$ are indicated by thin dot-dashed lines for the purple curve. Also shown for comparison is the integrated light curve produced by equation~\ref{eq:L_conv_t_planar_non_inst} (thin dash-dotted purple line).  \textit{Inset:} The initial shape of the shock for each model, at the time when shock acceleration begins (note that this differs from what is plotted in the inset of Figure~\ref{fig:Lightcurve_NonInstPlaneParallel}).  For the $\epsilon=0.2$ case (purple), the angle $\theta_{\rm obq}$ where the breakout first becomes oblique is indicated by a dotted purple line.
}
\label{fig:Lightcurve_NonInstOblique}
\end{figure}

Putting it all together, the light curve in the non-instantaneous oblique regime is described by
\begin{equation}
\label{eq:lightcurves_oblique}
    L(t) \approx
    \begin{cases}
    \Lbo \theta_{\rm bo}^2, & t<\tbo \\
    \Lbo \theta_{\rm bo}^2 (t/\tbo)^{(1-k)/(1+k)}, & \tbo < t < t_{\rm obq} \\
    L_{\rm asph}(t), & t_{\rm obq} < t < t_{\rm eq} \\
    L_{\rm sph}(t), & t > t_{\rm eq}\,.
    \end{cases}
\end{equation}
In Figure \ref{fig:Lightcurve_NonInstOblique}, we show the predicted light curves of aspherical explosions in this regime, for several choices of the asphericity parameter $\epsilon$ (as before, $\vstar/\vbo=0.01$ is adopted for illustrative purposes).  As expected, in the oblique breakout regime the duration of the initial flash becomes shorter as the asphericity is increased, since for high asphericity, the mushroom cloud which blocks the view of the breakout forms earlier.  We again compare the accurate light curve obtained from equation~\ref{eq:L_conv_t_planar_non_inst} (thin dot-dashed purple line) to the crude estimate given by equation~\ref{eq:lightcurves_oblique}.  Note that in this case, both the integrated light curve and the approximate one cut off sharply as soon as the breakout becomes oblique.

\subsection{Light travel time, viewing angle, and other caveats}
\label{sec:lighttravel}

So far, we have only considered the emission in the case where the observer is located along the symmetry axis, and where the diffusion time of the breakout layer satisfies $\tbo > R_*/c$ so that the light-crossing time is unimportant.  In this section, we consider how the breakout is affected when these assumptions are relaxed, and discuss a few other caveats that could affect the observed emission.

We first address how the light curves are altered when viewed by an observer located at an angle of $\theta_{\rm obs}$ from the axis.  For now we retain the assumption $\tbo > R_*/c$; the $R_*/c > \tbo$ case will be treated below.  Our goal is simply to give some idea of the extent to which the viewing angle can change the observed emission.   Therefore, for simplicity, we only compare the light curves for the cases where the observer is on the axis ($\theta_{\rm obs}=0$) or in the equatorial plane ($\theta_{\rm obs}=\upi/2)$; the emission for an arbitrary viewing angle will always be intermediate to these two extremes.  Full treatment of the emission as a function of the viewing angle is left to future work.

The viewing angle $\theta_{\rm obs}$ only affects the light curve in the case of a non-instantaneous breakout, since in an instantaneous breakout, the emission is the same as for a spherically symmetric breakout, which is identical regardless of viewing angle.  In the non-instantaneous case, differences between the early light curves for on-axis and off-axis observers arise primarily due to the fact that for a given surface element, the two observers see different projected areas.  In general, a surface element with area ${\rm d}A$, located at angle $\theta$, has a projected area ${\rm d}A_{\rm proj}={\rm d}A \, {\cos(\theta_{\rm obs}-\theta)}$ when viewed from an angle of $\theta_{\rm obs}$.  Assuming the surface is Lambertian (i.e., it has constant brightness per unit projected area, as expected for a blackbody), the luminosity of a given surface element is proportional to ${\rm d}A_{\rm proj}$.  Thus, if the shock sweeps out a projected area ${\rm d}A_{\rm proj}$ in time ${\rm d}t$, the observed luminosity is $L\propto {\rm d}A_{\rm proj}/{\rm d}t$.  

For an on-axis observer with $\theta_{\rm obs}=0$, ${\rm d}A_{\rm proj}={\rm d}A \cos \theta$. For small angles $\theta \ll 1$, ${\rm d}A_{\rm proj}\approx {\rm d}A$; therefore, the light curve at early times, which is produced by material close to the axis, is the same as what we have computed in Sections~\ref{sec:instantaneous}--\ref{sec:quasioblique}.  At later times, the emission from higher latitudes is damped by a factor of $\cos \theta$.  This makes the breakout somewhat fainter than our prediction at the end of the initial plateau phase, and smooths the transition from the initial breakout flash to the later adiabatically cooling phase (compare, e.g., the solid and dot-dashed lines in Figures~\ref{fig:Lightcurve_NonInstPlaneParallel} and~\ref{fig:Lightcurve_NonInstOblique}). However, other than that, the light curve is not significantly altered, and the behaviour is qualitatively the same as what we have previously shown.

The situation changes for an observer at the equator.  In this case, $\theta_{\rm obs}=\upi/2$, so  ${\rm d}A_{\rm proj}={\rm d}A \sin\theta$, and emission from near the poles is damped by a factor of $\sin \theta$.  Therefore the initial observed luminosity, which is produced in the region where $\theta< \theta_{\rm bo} \ll 1$, is reduced by a factor of $\approx \theta_{\rm bo}$.  Furthermore, the luminosity scales differently with time: in time $t$, the shock sweeps out an area of $\sim R_*^2 \theta^2$, but a \textit{projected} area of only $\sim R_*^2 \theta^3$, such that the luminosity evolves with time as $L \propto \theta^3/t \propto t^{(2-k)/(1+k)}$.  Note that for $1<k<2$, an on-axis observer sees a falling light curve during the prompt phase, while an observer at the equator sees a rising light curve.

The emission during the adiabatically cooling phase in our model is approximately spherically symmetric, and is not significantly affected by viewing angle, regardless of the relevant breakout regime.

The same procedure used to derive the integral expression~\ref{eq:L_conv_t_planar_non_inst} for an on-axis observer can be followed to derive a similar expression for an observer at the equator.  The only difference is the projected area of the annuli.  When viewed from the equator, the region between $\theta$ and $\theta + {\rm d}\theta$ has a projected area of $4 R_*^2 \sin^2 \theta {\rm d}\theta$, where we have taken into account the fact that an equatorial observer sees emission from both hemispheres.  Then, following the same steps as described above equation~\ref{eq:L_conv_t_planar_non_inst}, we obtain
\begin{equation}
    \label{eq:L_conv_equator}
    L(t)=\frac{4}{\upi} \int_0^{\upi/2} L_{\rm sph}(t-t_0(\theta)) \sin^2\theta \, {\rm d}\theta,
\end{equation}
where as before $t_0(\theta)$ is given by equation~\ref{ttheta}.

\begin{figure*}
\centering
\includegraphics[width=\textwidth]{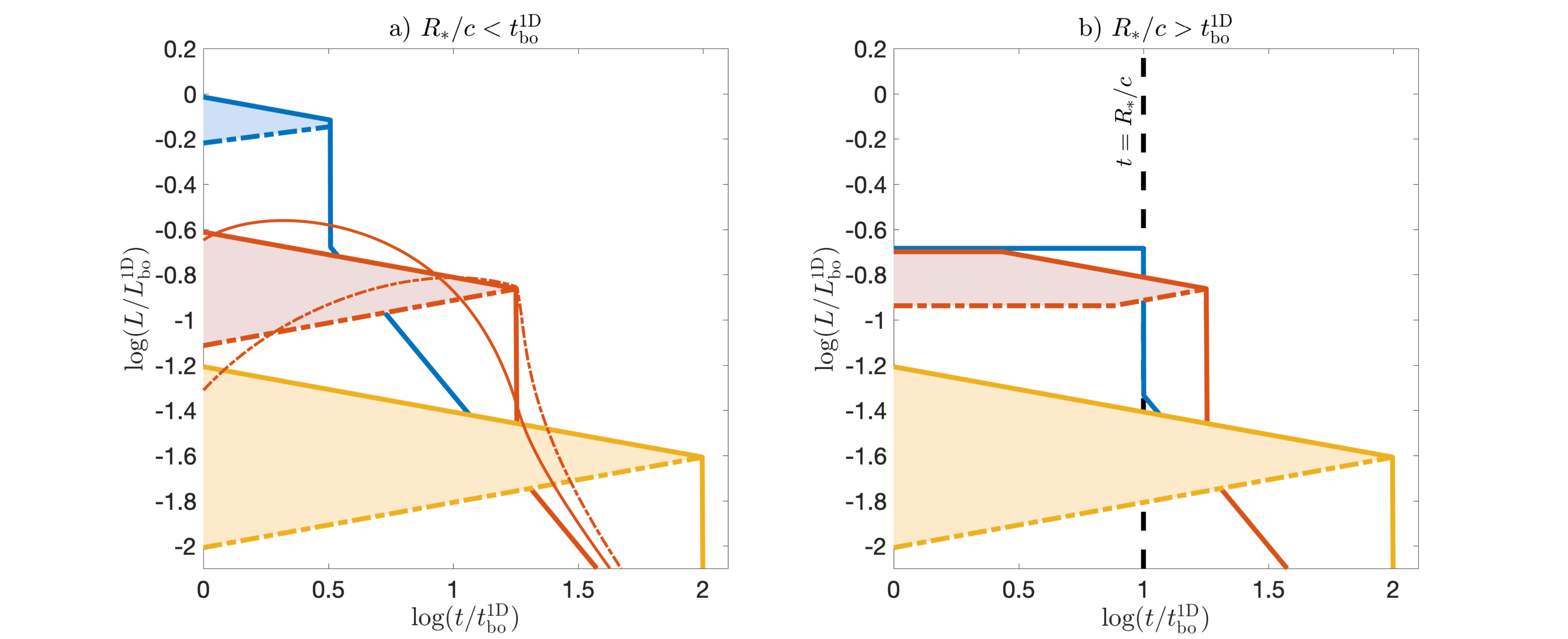}
\caption{Comparison between the observed light curves when the diffusion time of the breakout layer, $\tbo$, is longer than the stellar light crossing time $R_*/c$ (panel a), or shorter than $R_*/c$ (panel b).  We assume $\vstar/\vbo=0.01$ and $k=1.5$ in both cases, and in the right panel set $R_*/c=10\tbo$.    For each panel, we show light curves for three values of $\epsilon$: $1\times 10^{-5}$ (blue), $5.6 \times 10^{-5}$ (red), and $3.1 \times 10^{-4}$ (yellow).  The luminosity measured by an on-axis observer is plotted using solid lines, while the luminosity for an observer in the equatorial plane is shown using dot-dashed lines. The observed luminosity from any viewing angle will lie within the coloured bands bounded by these lines. In panel a), the more accurate light curves calculated via the integrals in equations~\ref{eq:L_conv_t_planar_non_inst} and~\ref{eq:L_conv_equator} are also shown for the $\epsilon=5.6\times10^{-5}$ case as thin red lines.  The time $t=R_*/c$ is also shown with a vertical dashed line in panel b).   The values of $\epsilon$ are specifically chosen to illustrate the ways in which the light curves in the $\tbo>R_*/c$ case are modified when $R_*/c>\tbo$ instead, with each of the three possibilities described in the text appearing in the right panel.  In the low-$\epsilon$ case (blue), all observers see the same light curve, which is equivalent to the spherically symmetric one.  In the intermediate-$\epsilon$ case (red), the light curve is modified in different ways for different viewing angles.  In the high-$\epsilon$ case, the light curve is not modified.}
\label{fig:OffAxisLightcurves}
\end{figure*}

The different light curves for $\theta_{\rm obs}=0$ and $\theta_{\rm obs}=\upi/2$ for the case where $\tbo>R_*/c$ are shown in panel a) of  Figure~\ref{fig:OffAxisLightcurves}, for a few representative values of $\epsilon$.  We point out that   the total energy of the observed breakout flash is always the same, for any viewing angle. The integrated light curves calculated using equations~\ref{eq:L_conv_t_planar_non_inst} and~\ref{eq:L_conv_equator} are also shown for comparison as thin lines for the $\epsilon=5.6\times10^{-5}$ case.  Whereas for an on-axis observer (solid lines) the integrated light curve agrees with the simple prediction at $t=\tbo$, and deviates from it once $t$ approaches $\twrap$, for an equatorial observer (dot-dashed lines) the opposite is true (i.e., the two curves agree at $t=\twrap$, but deviate from one another at $t=\tbo$).  This occurs because for on-axis observers, projected area effects are most important at the \textit{end} of the breakout when $\upi/2-\theta \ll 1$, while for observers at the equator projected area effects are important at the \textit{beginning} of the breakout, when $\theta \ll 1$.

We now cover the case where the light crossing time of the star exceeds the diffusion time of the breakout layer.  Panel b) of Figure~\ref{fig:OffAxisLightcurves} shows the modifications to the light curve when $R_*/c > \tbo$.  In this case, there are two fundamental differences as compared to the $\tbo> R_*/c$ case.  The first difference is relatively straightforward: because the duration of a spherical breakout is now set by $R_*/c$, the condition for an instantaneous breakout is now $\twrap<R_*/c$, rather than $\twrap<\tbo$.  Plugging this condition into equation~\ref{twrap}, we find that the transition between the instantaneous and non-instantaneous regimes occurs at $\epsilon=v_*/c$, instead of $\epsilon=\tbo/\tstar$ as before.  For any breakout with $\epsilon < v_*/c$, the shock crosses the star in less than a light-crossing time, so the initial emission is smeared over $R_*/c$, and as discussed in Section~\ref{sec:instantaneous}, the initial luminosity is $L\approx 2\Lbo (c\tbo/R_*)$ (as depicted by the blue curve in Figure~\ref{fig:OffAxisLightcurves}).

The second, more subtle difference occurs when the breakout is non-instantaneous.  As discussed in Section~\ref{sec:breakoutclasses}, in the non-instantaneous case the emission coming from a spherical cap with angular size $\theta_{\rm bo} < 1$ is smeared out by diffusion.  When light crossing time is important, there is similarly a region extending out to an angle $\theta_{\rm lc}<1$ over which the emission is smeared by light travel time.  The value of $\theta_{\rm lc}$ is found by locating the point where the component of the pattern velocity along the line of sight equals the speed of light, or equivalently by setting the light crossing time of the region, $t_{\rm lc}$, equal to the time for the breakout to traverse the region.  Because the light crossing time of a spherical cap depends on viewing angle, $\theta_{\rm lc}$ is a function of $\theta_{\rm obs}$, which makes this case significantly more complicated than the diffusion-dominated case.  For $\theta_{\rm obs}=0$ ,  the light crossing time of the cap where $\theta<\theta_{\rm lc}$ is $t_{\rm lc} \approx R_*\theta_{\rm lc}^2/2c$; then, setting $t_{\rm lc}=t(\theta_{\rm lc})$ in equation~\ref{ttheta}, we find $\theta_{\rm lc} \approx (\vstar/\epsilon c)^{1/(k-1)}$, and therefore $t_{\rm lc} \approx (R_*/c)(\vstar/\epsilon c)^{2/(k-1)}$.  On the other hand, if $\theta_{\rm obs}=\upi/2$, we have $t_{\rm lc}\approx R_*\theta_{\rm lc}/c$, and obtain $\theta_{\rm lc}\approx (\vstar/\epsilon c)^{1/k}$ and $t_{\rm lc}\approx(R_*/c)(\vstar/\epsilon c)^{1/k}$.   Here we have dropped order-unity prefactors for simplicity, in order to ensure that the light curve segments in Figure~\ref{fig:OffAxisLightcurves} connect smoothly.  

Depending on how $t_{\rm lc}$ compares with $\tbo$, there are two possible behaviours.  First, if $t_{\rm lc}>\tbo$, the duration of the initial constant-luminosity part of the light curve is $\approx t_{\rm lc}$, and the on-axis luminosity is $\approx \Lbo \theta_{\rm lc}^2$. When viewed from the equator, the initial luminosity is lower by an additional factor of $\theta_{\rm lc}$ due to the projected area considerations discussed above.  This situation is depicted by the red lines in panel b) of Figure~\ref{fig:OffAxisLightcurves}.   On the other hand, if $t_{\rm lc}<\tbo$, then the region where light travel is important is sufficiently small, so that the initial luminosity is set by the diffusion time of the breakout layer.  The light curve in this case is the same as it would be in the $\tbo < R_*/c$ case (as shown by the yellow lines in panel b) of Figure~\ref{fig:OffAxisLightcurves}).  The transition between the light travel time-dominated and diffusion time-dominated regimes occurs at a critical value of $\epsilon$, $\epsilon_{\rm crit}$, which depends on $\theta_{\rm obs}$, and for which $t_{\rm lc}=\tbo$.  Considering the cases of $\theta_{\rm obs}=0$ and $\theta_{\rm obs}=\upi/2$, we find that $\epsilon_{\rm crit}$ lies in the range $(\vstar/c)(R_*/c\tbo)^{(k-1)/2} <  \epsilon_{\rm crit} < (\vstar/c)(R_*/c\tbo)^{k}$.  

To give a physical example, for a Wolf--Rayet star, typically $\vstar/c \approx 0.01$ while $R_*/c\tbo \approx 100$.  In this case, assuming $k=1$,  $\epsilon_{\rm crit}\approx 0.01$ for an on-axis observer, but $\epsilon_{\rm crit} \approx 1$ for an equatorial observer.  Therefore, in WR stars, light travel time is expected to alter the  light curve for on-axis observers only if the asphericity is small ($\epsilon \la 0.01$), but for equatorial observers the light curve will be affected even for large asphericity ($\epsilon \la 1$).  For supergiant stars, which have $\vstar/c \approx 0.01$ and $R_*/c\tbo \la 5$, light travel time is generally only important for slightly aspherical explosions with $\epsilon \la 0.01-0.1$.

To sum up, the situation when $R_*/c>\tbo$ is as follows.  For $\epsilon < \vstar/c$, light travel time is important across the whole surface, and the light curve is identical to the usual spherically symmetric one.  For $\vstar/c < \epsilon < \epsilon_{\rm crit}$, only the emission from within a small region close to the axis is impacted by light travel time, and the early light curve is modified accordingly.  For $\epsilon > \epsilon_{\rm crit}$, light travel time does not affect the emission from any part of the surface, and the light curve is no different than in the $\tbo>R_*/c$ case.  (Note that if the observer is on-axis and $k<1$, then $\epsilon_{\rm crit}< \vstar/c$, so the situation is slightly different. In this case light travel time is non-negligible over the entire surface for $\epsilon<\epsilon_{\rm crit}$; important \textit{except for} a small region near the axis for $\epsilon_{\rm crit}<\epsilon<\vstar/c$; and not relevant at all for $\epsilon > \vstar/c$.) The full piecewise expressions for the light curves in each of these scenarios are given in Appendix~\ref{appendixc}. 

The light curves predicted by our model can be compared to numerical light curves calculated by, e.g., \citet{suzuki16} (see their Figure 7) and \citet{afsariardchi18} (their Figure 13).  In the case of \citet{suzuki16}, a direct comparison of their parameters and ours is difficult because they parametrize the explosion asphericity deep within the star at the inner boundary of the simulation, instead of near the edge of the star where shock acceleration begins. Based on their Figure 3, it appears that by the time the shock breaches the stellar surface, the degree of asphericity is relatively small.  Judging by the appearance of the light curves in their Figure 7, their $a=0.2$ and $a=0.5$ cases are likely in the non-instantaneous plane-parallel regime, while their $a=0.8$ case may be marginally in the non-instantaneous oblique regime. In spite of the different setup, their results are qualitatively similar to what we have predicted here: the aspherical breakout emission is fainter and longer-lived than in the spherical case, and is characterized in the on-axis case by an initial plateau, followed by a rapid drop to rejoin the spherically symmetric light curve.  In addition, they find a falling light curve for on-axis observers, and a rising light curve for off-axis observers, which is consistent with our model for $1< k < 2$.  The off-axis emission also stays constant for longer before rising, which appears similar to our non-instantaneous light travel time-modified case (i.e., the red curves in panel b) of our Figure~\ref{fig:OffAxisLightcurves}).

\citet{afsariardchi18}, on the other hand, consider significantly more aspherical explosions.  They use a parameter $\epsilon_{\rm AM}$\footnote{In \citet{afsariardchi18}, this parameter is simply called $\epsilon$.  We have added the subscript `AM' to distinguish it from our parameter $\epsilon$ defined in Section~\ref{sec:breakoutclasses}.} to characterize the asphericity.
Their fiducial $\epsilon_{\rm AM}=0.26$ case corresponds to a scenario where the time for the shock to wrap around the star is comparable to $\tstar$, and therefore it is likely in the highly aspherical regime which we do not consider here (as can be seen in their Figure 2b, the shocked region for this model is rather narrow, with an aspect ratio of $\ga 2$ at the time of breakout from the poles).  We therefore focus on their intermediate $\epsilon_{\rm AM}=0.09$ case.  They consider three different progenitors.  For their red supergiant and blue supergiant models, the breakout is in our non-instantaneous planar regime.  Their results in this case are similar to those of \citet{suzuki16}, and again are qualitatively consistent with our model.  Their Type Ic model exhibits non-radial flow and is likely in our non-instantaneous oblique breakout regime.  The features of this model---an initial luminosity which is fainter than in the spherical case, an early sharp drop, and a luminosity that declines more slowly than the spherical prediction---are all consistent with the predictions of our breakout model in the case of an on-axis observer.

One feature of the light curve observed by both \citet{suzuki16} and \citet{afsariardchi18} which is not replicated by our model is the appearance of a second peak in the light curve for equatorial observers.  This discrepancy is purely due to a difference in the assumed shape of the shock.  As discussed earlier in Section~\ref{sec:breakoutclasses}, we assume an idealized shock shape for which the inclination angle $\psi_0$ monotonically increases with $\theta$ (equation~\ref{psiansatz}).  In contrast, both \citet{suzuki16} and \citet{afsariardchi18} use a shock shape for which $\psi_0\rightarrow 0$ as $\theta \rightarrow \upi/2$.  In this case, the pattern speed $v_{\rm p} \propto (\sin \psi_0)^{-1}$ reaches a minimum somewhere along the stellar surface, and then climbs again to $v_{\rm p} \rightarrow \infty$ as $\theta \rightarrow \upi/2$.  Consequently, in their models, the energy in the equatorial region is released over a much shorter time-scale than in our model, which causes the luminosity to rebrighten to a second peak at $t \sim \twrap$.  As discussed by \citet{afsariardchi18}, the second peak only occurs for sufficiently large viewing angles (as otherwise it is damped due to the small projected area of the equatorial region), and only when the breakout is parallel across the whole surface (since if the breakout becomes oblique, the sideways spray of ejecta blocks the view of the breakout for all $\theta>\theta_{\rm obq}$).  If we were to adopt a more sophisticated functional form for $\psi_0$, with $\psi_0=0$ at $\theta=\upi/2$, we would also expect to see this secondary peak in the case of a non-instantaneous parallel breakout viewed from the equator.

To conclude this section, we briefly consider how some of our other simplifying assumptions may impact the light curve.  First, in calculating the adiabatically cooling emission, we assumed that the radiation diffuses out along radial rays, so that the emission is basically isotropic.  However, particularly for very oblique breakouts, an appreciable fraction of the radiation likely escapes in non-radial directions.  Thus, the observed emission during the adiabatically cooling phase may also depend somewhat on viewing angle.  In addition, we stress that we have only considered the bolometric output of the breakout; the emission observed in a given waveband may be markedly different.

One other possibility that we have not considered is additional dissipation of energy due to the collision of ejecta in the equatorial plane outside the star \citep[as discussed by, e.g.,][]{matzner13,afsariardchi18}.  In an oblique breakout, the ejecta have significant non-radial velocities, and some ejecta even have velocities directed downward towards the equatorial plane.  These ejecta will eventually reach the equator where, due to the reflective symmetry of the system, they will encounter ejecta from the opposite side of the star.  The ensuing collision will dissipate some of the kinetic energy into internal energy which is then radiated.  This effect is expected to produce additional transient emission on a time-scale of $\sim$ a few $R_*/\vbo$ \citep[e.g.,][]{matzner13}. It is not clear how bright this emission will be, but it may alter our estimated light curve at times around $\sim \ts$.

\section{Astrophysical Implications}\label{sec:implications}

Our key finding that sufficiently aspherical shock breakout produces a longer but fainter prompt breakout flash has important implications for observational searches of shock breakout phenomena.  As discussed in the introduction, the limiting observational factor in detecting shock breakout is not typically the lack of sensitivity, but rather the difficulty of probing short time-scales.  Therefore, although the signal is fainter, we expect that the prolonged time-scale will none the less make it easier to detect.   Moreover, in the non-instantaneous plane-parallel breakout regime, although the average luminosity of the breakout flash is fainter than in a spherical breakout, the luminosity towards the end of the flash actually exceeds the predicted luminosity from the spherical model at that time (compare, e.g., the blue, red and yellow curves to the black curve in Figure~\ref{fig:Lightcurve_NonInstPlaneParallel}).  The reason is that, whereas in the spherical model adiabatic cooling sets in immediately for all the ejecta, in the non-instantaneous breakout case the onset of adiabatic cooling for the high-latitude material (where most of the mass and energy is located) is delayed until the breakout crosses to the equator.  Since the tail end of the breakout is the part most likely to be caught in observations, this further strengthens our conclusions about improved detectability.

\begin{figure*}
    \centering
    \includegraphics[width=\textwidth]{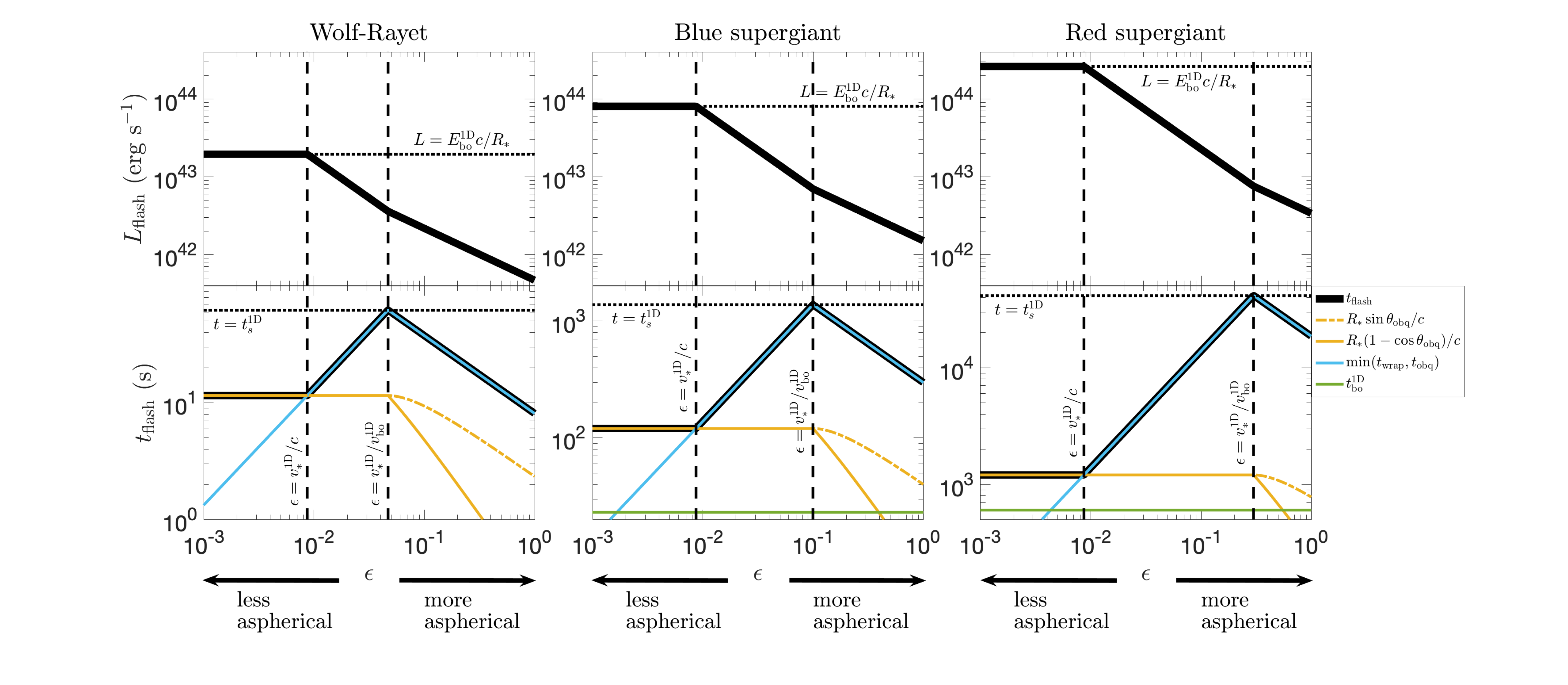}
    \caption{Comparison of the duration $t_{\rm flash}$ and bolometric luminosity $L_{\rm flash}=L(t_{\rm flash})$  of the initial breakout flash as functions of the asphericity parameter $\epsilon$, for three different progenitors: a Wolf--Rayet (WR) star with radius $5 R_\odot$ (left), a blue supergiant (BSG) with radius $50 R_\odot$ (middle), and a red supergiant with radius $500 R_\odot$ (right).  In all cases, we adopt an explosion energy $E=10^{51}\,$ergs and a stellar mass of $M_*=15M_\odot$, and set $k=1.5$. We assume an opacity of $\kappa=0.34\,\rm{cm}^2\,\rm{g}^{-1}$ for the RSG and BSG cases, and $\kappa=0.2\,\rm{cm}^2\,\rm{g}^{-1}$ for the hydrogen-free WR star case.  The outer density profile is described by $n=3$ for the WR and BSG, and $n=1.5$ for the RSG, as appropriate for radiation-dominated and convection-dominated energy transport, respectively.  In all cases we find that $R_*/c>\tbo$, although the RSG case is marginal.  The vertical dashed lines indicate the transition between instantaneous and non-instantaneous breakout at $\epsilon=\vstar/c$, and the transition between parallel and oblique breakout at $\epsilon=\vstar/\vbo.$  \textit{Upper panels:} The luminosity of the breakout flash (heavy black line), as compared with the luminosity of a spherical breakout (thin dotted line), which is $\approx \Ebo c/R_*$.  \textit{Lower panels:} The duration of the breakout flash (heavy black line), showing the maximum possible duration of $t_{\rm flash}=R_*/\vbo=\ts$ at $\epsilon=\vstar/\vbo$ as a dotted black line.  Note that the vertical axis is scaled differently in each case.  Also shown are: the light crossing time of the parallel breakout region with angular size $\theta_{\rm obq}$ (yellow), which is $(R_*/c)(1-\cos\theta_{\rm obq})$ for an on-axis observer (solid) or $(R_*/c)\sin\theta_{\rm obq}$ for an observer in the equatorial plane; the shock crossing time of the parallel breakout region (cyan), which is $\twrap$ for a parallel breakout or $t_{\rm obq}$ for an oblique breakout; and the diffusion time of the breakout layer, $\tbo$ (green), which is too small to show for the WR star case.  For a given $\epsilon$, the flash duration is set by the longest of these time-scales. }
    \label{fig:BreakoutDuration}
\end{figure*}

To give a better sense of how asphericity may impact the observed properties of the initial breakout pulse, we apply our model to three typical supernova progenitors: a Wolf--Rayet (WR) star of radius $5 R_\odot$, a blue supergiant (BSG) of radius $50 R_\odot$, and a red supergiant (RSG) of radius $500 R_*$.  For each case, we calculate the observed duration $t_{\rm flash}$ and bolometric luminosity $L_{\rm flash}=L(t_{\rm flash})$ as functions of the asphericity parameter $\epsilon$, assuming an explosion energy of $10^{51}\,\rm{ergs}$ and a stellar mass of $15 M_\odot$ 
The results are shown in Figure~\ref{fig:BreakoutDuration}.  As $\epsilon$ goes from 0 to 1, we find that $L_{\rm flash}$ is decreased by up to 2 orders of magnitude compared to a spherically symmetric explosion.  
Meanwhile, the pulse duration is extended up to a maximum value of $\sim 1$ minute for a WR star, $\sim$ half an hour for a BSG, and $\sim 10$ hours for an RSG.  The larger the progenitor, the more asphericity is needed to maximize $t_{\rm flash}$, with the optimal value of $\epsilon$ ranging from 0.06 for the WR case to 0.3 for the RSG case.  

However, even if the prompt breakout emission is detected, distinguishing between the different breakout scenarios may be difficult.  The prompt breakout flash alone is not sufficient to differentiate between the spherical, non-instantaneous parallel, and non-instantaneous oblique regimes, because the appearance of the initial plateau is similar across regimes.  A long-duration breakout transient could be due to asphericity, but it could equally well be caused by the progenitor simply having a large radius.  To make matters worse, there is a degeneracy between the two non-instantaneous cases.  In the plane-parallel limit, the breakout energy is $\Ebo \propto R_*^2$ and the duration is $t_{\rm flash} \approx R_*/v_{\rm p}(\theta=\upi/2) \propto R_*$, so the luminosity scales as $L_{\rm flash} \approx \Ebo/t_{\rm flash} \propto R_*$.  On the other hand, in the oblique case the energy is $\Ebo \theta_{\rm obq}^2 \propto R_*^2 \theta_{\rm obq}^2$, the duration is $t_{\rm flash} \approx R_*\theta_{\rm obq}/\vbo \propto R_*\theta_{\rm obq}$, and the luminosity is $L_{\rm flash} \approx \Ebo/t_{\rm flash} \propto R_* \theta_{\rm obq}$.  Therefore, the breakout flash produced by an oblique breakout with a radius $R_*$ and a transition angle $\theta_{\rm obq}$ is nearly identical to the flash produced by a non-instantaneous parallel breakout with a radius $R_* \theta_{\rm obq}$.

Given these difficulties, is there a way to easily differentiate between parallel and oblique breakout signatures?  It may be possible if the signal can still be tracked  after the initial breakout flash.  Before investigating this, we first stress again that 1) our model only considers the bolometric light output, which may not be accessible from observations and 2) robust detection of shock breakout will be difficult in any case, owing to the short time-scales involved.  None the less, let us suppose a very optimistic scenario where far UV or X-ray observations are available so that we have some handle on the bolometric luminosity, and where time-scales of order $\la R_*/\vbo$ are able to be probed.  In this best-case scenario, what would the light curves look like in each breakout regime?  Here we adopt nominal values of $k=1$, $n=3$, and $\mu=0.19$ for concreteness. In the spherical or parallel non-instantaneous case, we might hope to catch the end of the initial pulse, which would initially appear as a flat plateau or shoulder, and then quickly drop off.  Afterwards, we would expect to see a broken power-law with an initial decay as $L \propto t^{-4/3}$ for $t<R_*/\vbo$, followed by  a break to a shallower decay $L \propto t^{-0.35}$ at $t>R_*/\vbo$.   On the other hand, the non-instantaneous oblique case would similarly start out as a flat plateau, but then rapidly drops off into a $L \propto t^{-0.15}$ decline instead.  In short, the parallel (oblique) cases can be identified according to whether the light curve declines as $t^{-4/3}$ ($t^{-0.15}$) shortly after the initial drop.   In principle, it should also be possible to distinguish an oblique breakout from a parallel one even if the initial flash is missed, by measuring the power-law slope of the light curve at late times.  However, for typical values the difference in the temporal indices is only $\approx 0.2$, so in practice the two scenarios may be difficult to robustly tell apart at times $\ga R_*/\vbo$.

A particularly interesting feature of our model is the existence of an upper limit on the breakout duration $t_{\rm flash}$ which can be obtained from aspherical breakouts.  The maximum duration is given by $t_{\rm flash} < R_*/\vbo$ (as shown in Figure~\ref{fig:BreakoutDuration}), where $\vbo$ is the maximum velocity to which ejecta are accelerated during the breakout. This condition can be used to test whether an event is consistent with an aspherical breakout origin: if the breakout duration and the photospheric velocity $v_{\rm phot}$ are measured, then an aspherical breakout must satifsy $t_{\rm flash}<R_*/v_{\rm phot}$, since $v_{\rm phot}$ places a lower limit on $\vbo$.  Here we apply this test to the supernova SN 2008D in order to explore whether its accompanying X-ray flash \citep[XRF 080109;][]{soderberg08} could plausibly have been produced by an aspherical explosion within a typical Wolf--Rayet star.  The duration of the X-ray flash in SN 2008D was $t_{\rm flash}\approx 490$ s, and the photospheric velocity at early times was $\approx 30,000 \, \text{km}\,\text{s}^{-1}$ \citep{mazzali08,modjaz09}.  For a typical Wolf--Rayet radius of $R_*=10^{11}\,\text{cm}$, we therefore find that $R_*/v_{\rm phot} \approx 30\,\text{s}$ and the condition is violated.  This suggests that asphericity alone is not sufficient to reproduce the long-duration X-ray flash; a larger breakout radius (or a combination of a larger radius and asphericity) is therefore needed.  Our result seems roughly consistent with the findings of \citet{couch11}, who applied numerical simulations of aspherical explosions in Wolf--Rayet stars to SN 2008D, but found that the simulated light curves fell off more quickly than the observations, resulting in under-luminous emission at late times (see, however, their discussion of non-LTE effects).

The upper limit on $t_{\rm flash}$ also has important consequences for low-luminosity GRBs ($ll$GRBs).  Here we focus on the most well-observed $ll$GRB, GRB 060218.  Several authors have suggested that the prompt X-ray emission in this event was powered by a mildly relativistic shock breakout in a dense circumstellar wind \citep[e.g.,][]{campana06,waxman07}, but an issue with this interpretation is that the inferred breakout radius, $r\approx 10^{12}\,\rm{cm}$, implies a duration of only $r/c \approx 300\,\rm{s}$, at odds with the observed duration of $\sim 3000\,\rm{s}$ \citep{campana06}.  Aspherical shock breakout has been suggested as a means to prolong the duration, thereby resolving this discrepancy \citep[e.g.,][]{waxman07}.  However, we have shown that asphericity only increases the duration by at most a factor of $(R_*/\vbo)/(R_*/c)=c/\vbo$.  Thus, for a mildly relativistic breakout with $\vbo \sim c$, the breakout duration is increased only by a factor of a few, if at all.  If GRB 060218 was produced by a relativistic shock breakout, a breakout radius considerably larger than $10^{12}\,\rm{cm}$ is therefore preferred \citep[see, however,][for arguments against a shock breakout origin]{ghisellini07,irwin16}. A larger breakout radius from an optically thick envelope extending to $\sim 10^{13}\,\rm{cm}$ is consistent with the prompt gamma-ray emission of this burst \citep[as shown by, e.g.,][]{nakar12,nakar15}, and has also been inferred from the early optical peak in the supernova SN 2006aj which accompanied the GRB \citep[e.g.,][]{nakar14,nakar15,irwin16}.

\section{Summary and conclusions}
\label{sec:summary}

We have considered a model for aspherical shock breakout in the simplified case of a bipolar shock which is extended along the symmetry axis.  We first used past results on oblique shock breakout to link the breakout properties at a given point on the stellar surface to the inclination angle of the shock when it first reaches the steep outer density gradient near the edge of the star and starts to accelerate.  Then, assuming a simple power-law form for the angular dependence of this initial inclination angle, $\psi \propto \epsilon \theta^k$, 
we calculated the breakout properties at each point along the surface, and examined how the overall breakout behaviour depends on the asymmetry parameters $\epsilon$ and $k$. For a prolate shock which first breaks out at the poles and then wraps around to the equator, the parameter $\epsilon$, defined in equation~\ref{epsilondef}, can be intuitively understood as the ratio $\epsilon \sim \twrap/\tstar$, where $\twrap$ is the difference in time between breakout at the poles and at the equator, and $\tstar\approx (R_*^2 M_*/E)^{1/2}$ is the time it takes the shock to cross the star (from the centre to the surface).  For spherical explosions, $\epsilon=0$, while more aspherical explosions have larger $\epsilon$.  As $\epsilon$ is increased, the shape of the shock becomes more narrow  (compare, e.g., the blue and orange curves in the right panel of Figure~\ref{fig:AsphericalShockParametrization}).

We identified four possible breakout scenarios, with the relevant case determined by comparing the quantity $\epsilon$ to quantities obtained from the spherically symmetric breakout theory (see table~\ref{table:regimes}).  Provided that $\epsilon$ is non-zero, the breakout starts at the poles and peels along the surface towards the equator.  For sufficiently small $\epsilon$, we find that the breakout is completed on a time-scale shorter than $\max(R_*/c,\tbo)$, where $\tbo$ is the diffusion time of the breakout layer in the spherically symmetric theory, and the resulting emission is effectively unchanged compared to the case of spherical symmetry (`effectively instantaneous breakout').  However, as $\epsilon$ increases, the speed at which the breakout moves across the surface decreases, and the duration of the breakout becomes set by the time for the breakout to wrap around the star (`non-instantaneous plane-parallel breakout').  The initial breakout pulse is prolonged in this case, and as the total energy released, $\Ebo$, remains unchanged, the emission is fainter.  

As $\epsilon$ is increased further, the shock inclination angle continues to steepen, eventually reaching the point that somewhere along the stellar surface the shock becomes highly oblique before its radiation can escape.  While there is always a neighborhood around the poles of angular size $\theta_{\rm obq}$ where obliquity is unimportant, for sufficiently large $\epsilon$ this region becomes small and most of the surface is in the oblique breakout regime (`non-instantaneous oblique breakout').  Once the breakout becomes oblique, we find that material is shunted to the side as it breaks out, forming an ejecta cloud which hides the remainder of breakout from the observer (as in the middle and right panels Figure~\ref{fig:PlanarVsObliqueBreakout}).  In this case, the observed breakout duration is the time for the breakout to travel to $\theta_{\rm obq}$, and the energy emitted during the initial pulse is only $\approx\Ebo \theta_{\rm obq}^2$; therefore, the breakout flash is both shorter and fainter compared to the plane-parallel case.  Finally, if $\epsilon$ is increased to values $\epsilon \ga 1$, then the shock actually becomes narrow, with an opening angle of $\la 1$ radian (`significantly aspherical breakout').  In this case the breakout is still oblique over most of the surface, but unlike the previous case, where the large shock inclination at breakout is induced by the sharp shock acceleration near the star's edge, in this case the shock already has a large inclination of $\sim 1$ radian even before it is accelerated. The behaviour in this case is more challenging to calculate, and is reserved for future work, although it is clear at least that the breakout duration will be even shorter than before.

For the first three scenarios, we estimated the bolometric light curve, considering both the shock breakout emission and the adiabatically cooling emission from the ejecta cloud thrown out during the breakout (see figures~\ref{fig:Lightcurve1D},~\ref{fig:Lightcurve_NonInstPlaneParallel},~\ref{fig:Lightcurve_NonInstOblique} and equations~\ref{lightcurvespherical},~\ref{lightcurves_planarnoninst},~\ref{lightcurves_obliquenoninst}).  The light curves are comprised of broken power-law segments with temporal indices depending on $k$, which governs how the initial shock inclination varies with angle (i.e., $\psi \propto \theta^k$), and $n$, which describes the density profile near the edge of the star (i.e., $\rho \propto (R_*-r)^n$).  A steep drop is observed in the non-instantaneous cases at the time when breakout ends, because the outermost layer of material ejected from high latitudes cools on a time-scale which is very short compared to the time when it is ejected.  In the non-instantaneous oblique case, we find that the luminosity at late times (which is produced by the cooling ejecta cloud), is fainter than the case of a spherical breakout, and declines more slowly.  This occurs because the maximum ejecta velocity is limited in an oblique breakout; for a given velocity $v_i$, only material within an angle $\theta_{v>v_i}<(\vstar/\epsilon v_i)^{1/k}$ of the axis is accelerated to speeds $> v_i$, where $\vstar \sim (E/M_*)^{1/2}$ is the bulk SN velocity.  The fastest ejecta (which are observed first) therefore have only a fraction of the energy they would have in the spherical case, resulting in an initially fainter light curve.  As the outflow expands, deeper and slower layers with larger $\theta_{v>v_i}$ are gradually revealed.  Eventually, once layers with velocity $v_i<v_*/\epsilon$ and $\theta_{v>v_i}  \sim 1$ become visible, the non-instantaneous oblique light curve rejoins the usual spherical one.

A sufficiently aspherical shock breakout produces a longer and fainter prompt flash compared with a spherical breakout. As the limiting factor in detecting shock breakout events is typically the difficulty in probing short time-scales rather than the lack of sensitivity, our results suggest that aspherical shock breakout will generally be easier to detect. Since the adiabatic cooling of high-latitude material is delayed by the time it takes the shock to wrap around the stellar surface, $\twrap$, the end of the plateau phase of a non-instantaneous breakout can outshine the emission of a spherically symmetric shock breakout at that time (e.g., Figure \ref{fig:Lightcurve_NonInstPlaneParallel}).  Interestingly, our model places an upper limit on the duration of the prompt breakout flash in an aspherical breakout.
As $\epsilon$ gradually increases from 0, the breakout duration is initially prolonged due to the increase in $\twrap$. However, when $\epsilon\ga \vstar/\vbo$, where $\vbo$ is the fastest velocity to which ejecta are accelerated, the breakout becomes oblique at an angle $\theta_{\rm obq}$, and the accompanying non-radial ejecta enshroud the stellar surface at high latitudes, blocking view of the breakout at times $t \ga R_*\theta_{\rm obq}/\vbo$.  The maximum possible duration, $R_*/\vbo$, occurs when $\epsilon \sim \vstar/\vbo$ and $\theta_{\rm obq}\sim 1$. 
We apply this result to show that both SN 2008D and GRB 060218 cannot be explained by an aspherical breakout from a standard Wolf--Rayet progenitor, owing to the long duration of the X-ray flashes associated with these events.

While the growing body of work on aspherical shock breakout in recent years has begun to shed light on this complex phenomenon, aspherical breakout still remains poorly understood compared to the spherical case. However, one thing is abundantly clear: even small departures from spherical symmetry can have considerable observational consequences, and therefore the field is ripe for future studies.  Some interesting theoretical avenues to explore include moving beyond the bolometric emission to calculate the spectrum and band-dependent light curves, and investigating the case of highly aspherical breakouts with initially narrow shocks, which may be formed by jets.  Meanwhile, future observations by missions such as ULTRASAT will be critical in shaping our understanding of the early light from stellar explosions, and may well detect aspherical shock breakout for the first time.

\section*{acknowledgments}
CI thanks T. Shigeyama, A. Suzuki, C. Omand, and K. Fujisawa for helpful discussions during a visit to RESCEU at the University of Tokyo in 2019.  IL acknowledges support from the Adams Fellowship.  This work was supported in part by the Zuckerman STEM Leadership Program (CI); by a consolidator ERC grant JetNS  and an ISF grant (EN); and by an advanced ERC grant TReX (TP). RS is supported by an ISF grant.

\section*{Data Availability}
The data underlying this article will be shared on reasonable request to the corresponding authors.

\appendix
\section{Glossary of symbols}
\label{appendixa}
In Table~\ref{table:symbols}, we define numerous important symbols which are used throughout the paper.

\begin{table*}
\centering
\caption{Glossary of symbols}
\begin{tabular}{l l}
\hline \hline
\multicolumn{2}{c}{Properties of the progenitor and explosion} \\ \hline 
$R_*$ & stellar radius \\
$M_*$ & stellar mass \\
$E$ & total explosion energy \\
$\alpha$ & power-law index of the star's inner density profile, $\rho(r) \propto r^{-\alpha}$ \\
$n$ & power-law index of the star's outer density profile, $\rho(r) \propto (R_*-r)^{-n}$ \\
$R_0$ & radius where the density profile becomes steep, and the shock begins to accelerate \\
\hline
\multicolumn{2}{c}{Relevant quantities from the spherically symmetric breakout theory}  \\
\hline
$\vstar$ & initial shock velocity at the onset of shock acceleration  \\
$\tstar$ & time for the shock to travel from the centre to the edge of the star, i.e. $\tstar \approx R_*/\vstar$ \\
$\vbo$ & shock velocity at the moment of breakout \\
$\ybo$ & distance from the stellar surface at which the breakout takes place, and also the initial width of the breakout layer \\
$\tbo$ & diffusion time through the breakout layer, and also its dynamical time \\
$\Ebo$ & energy contained in the breakout layer \\
$\Lbo$ & luminosity of the breakout flash \\
$\ts$ & time-scale for the outflow's radius to double following the breakout, i.e. $\ts \approx R_*/\vbo$ \\
\hline
\multicolumn{2}{c}{Parameters and derived quantities in the aspherical breakout model}  \\
\hline
$\theta$ & polar angle \\
$\psi_0(\theta)$ & initial inclination angle of the shock when it begins to accelerate at $r=R_0$ \\
$\psi_{\rm bo}(\theta)$ & inclination angle of the shock at the moment of breakout \\
$\epsilon$ & normalization of $\psi_0$; $\epsilon=\psi_0(\theta=\upi/2)$ \\
$k$ & power-law index describing $\psi_0$; $\psi_0(\theta) \propto \theta^{k}$ \\
$v_{\rm p}(\theta)$ & pattern velocity at which the breakout travels along the stellar surface \\
$\twrap$ & time it takes for the breakout to reach the equator \\
$\theta_{\rm bo}$ & angle over which the emission is smeared by diffusion \\
$\theta_{\rm obq}$ & angle at which the breakout transitions from parallel to oblique \\
$\theta_{\rm nr}$ & angle where the shock inclination angle exceeds $\sim 1$ radian \\
$t_{\rm obq}$ & time at which the breakout transitions from parallel to oblique \\
$t_{\rm eq}$ & time after which an oblique shock breakout has the same luminosity as a spherically symmetric breakout \\
\hline
\multicolumn{2}{c}{Properties of the outflow following breakout} \\
\hline
$i$ & index used to divide the outflow into discrete layers/shells \\
$v_i$ & initial velocity of the $i$-th shell \\
$\theta_i$ & initial angular size of the $i$-th shell \\
$t_i$ & time at which the $i$-th shell was ejected \\
$\rho_i$ & initial density of the $i$-th shell \\
$y_i$ & initial thickness of the $i$-th shell \\
$E_i$ & initial energy contained in the $i$-th shell \\
$\tau_i$ & initial optical depth of the $i$-th shell \\
$\tilde{v_i}(t)$ & velocity of the `luminosity shell' with optical depth satisfying $\tau=c/\tilde{v_i}$, which dominates the emission at time $t$ \\
$\tilde{\theta_i}(t)$ & initial angle of the luminosity shell when it was ejected \\
$\tilde{E_f}(t)$ & energy contained in the luminosity shell after accounting for adiabatic losses \\
\hline
\multicolumn{2}{c}{Observables related to the light curve} \\
\hline
$L_{\rm flash}$ & luminosity of the initial breakout flash \\
$t_{\rm flash}$ & duration of the initial breakout flash \\
\hline \hline
\end{tabular}
\label{table:symbols}
\end{table*}

\section{Time evolution of relevant quantities in a non-instantaneous oblique breakout}
\label{appendixb}

In Section~\ref{sec:quasioblique}, we simplified the exponents by assuming $n=3$ and $\mu=0.19$.  The full expressions for arbitrary $\mu$ and $n$ are as follows:

\begin{equation}
    t_i \propto \theta_i^{1+k} \,
\end{equation}
\begin{equation} 
    v_i \propto \theta_i^{-k} \,
\end{equation}
\begin{equation}
    \rho_i \propto \theta_i^{k/\mu} \,
\end{equation}
\begin{equation} 
    y_i\propto \theta_i^{k/(\mu n)} \,
\end{equation}
\begin{equation} 
    m_i\propto \theta_i^{(k(n+1)+2\mu n)/(\mu n)} \,
\end{equation}
\begin{equation}
    \tau_i\propto \theta_i^{k(n+1)/(\mu n)} \,
\end{equation}
\begin{equation} 
    E_i\propto \theta_i^{(k(n+1-2\mu n)+2\mu n)/(\mu n)} \,
\end{equation}
\begin{equation} 
    \Tilde{\theta_i} \propto t^{2\mu n/(k(1+n+\mu n)+2\mu n)} \,
\end{equation}
\begin{eqnarray}
    \Tilde{E}_f & \propto & t^{[2(k(n+1-2\mu n)+2\mu n) + (\gamma-1)(3\mu n k-3nk-k-2\mu n)]/[k(1+n+\mu n) + 2\mu n]} \, \\
    \nonumber & \propto & t^{[k(5+3n-9\mu n)+10\mu n]/[3k(1+n+\mu n) + 6\mu n]}\,, \text{ for } \gamma=4/3 
\end{eqnarray}

\section{Light curves modified by light travel time and/or viewing angle} \label{appendixc}

As discussed in Section~\ref{sec:lighttravel}, if the light--crossing time of the star $R_*/c$ is longer than the diffusion time of the breakout layer $\tbo$, then light travel time effects may alter the signal over some or all of the stellar surface.  Depending on the value of the asphericity parameter $\epsilon$, there are three possible behaviours.  If $\epsilon < \vstar/c$, then any emission from times $t<R_*/c$ is smeared over the light--crossing time of the star.  If $\vstar/c < \epsilon < \epsilon_{\rm crit}$, then only the emission from a spherical cap of angular size $\theta_{\rm lc}$ is affected by light travel time, with the emission from that cap being smeared over the light crossing time of the cap, $t_{\rm lc}$.  Finally, if $\epsilon>\epsilon_{\rm crit}$, then light crossing time is not important.  The values of $\theta_{\rm lc}$, $t_{\rm lc}$, and $\epsilon_{\rm crit}$ depend on the observer's viewing angle relative to the axis, $\theta_{\rm obs}$.

Here, we provide the segmented power-law light curves in each of these three scenarios.  In each case, we give the light curve for both an on--axis observer ($\theta_{\rm obs}=0$), and an observer located in the equatorial plane ($\theta_{\rm obs}=\upi/2$). (Note that in the $\epsilon < \vstar/c$ case, the light curve is the same regardless of $\theta_{\rm obs}$.)  For these values of $\theta_{\rm obs}$, the relevant values of $\theta_{\rm lc}$, $t_{\rm lc}$, and $\epsilon_{\rm crit}$ are:
\begin{equation}
\theta_{\rm lc} \approx \begin{cases}
(\vstar/\epsilon c)^{1/(k-1)}, & \theta_{\rm obs}=0 \\
(\vstar/\epsilon c)^{1/k}, & \theta_{\rm obs}=\upi/2
\end{cases}    ,
\end{equation}
\begin{equation}
t_{\rm lc} \approx \begin{cases}
(R_*/c)(\vstar/\epsilon c)^{2/(k-1)}, & \theta_{\rm obs}=0 \\
(R_*/c)(\vstar/\epsilon c)^{1/k}, & \theta_{\rm obs}=\upi/2
\end{cases}    ,
\end{equation}
and
\begin{equation}
\epsilon_{\rm crit} \approx \begin{cases}
(\vstar/c)(R_*/c \tbo)^{1/(k-1)}, & \theta_{\rm obs}=0 \\
(\vstar/c)(R_*/c \tbo)^{k}, & \theta_{\rm obs}=\upi/2
\end{cases}   .
\end{equation}

\subsection{Instantaneous breakout ($\epsilon<\vstar/c$)}
The light curve is the same as in the instantaneous, spherical case (Section~\ref{sec:instantaneous}, except that the duration of the initial pulse is extended up to $R_*/c$, with a characteristic luminosity $\Ebo c/R_*$:
\begin{equation}
    L(t) \approx \begin{cases}
    2\Lbo(c\tbo/R_*), & t< R_*/c \\
    L_{\rm sph}, & t>R_*/c
    \end{cases},
\end{equation}
where $L_{\rm sph}$ is given by equation~\ref{lightcurvespherical}.

\subsection{Non-instantaneous breakout, light travel time is important over part of the surface ($\vstar/c<\epsilon<\epsilon_{\rm crit}$)}
For non-instantaneous breakouts, light travel time and viewing angle only affect the light curve during the initial breakout pulse, i.e. for times $t<\min(\twrap,t_{\rm obq})$.  At later times, during the adiabatically cooling phase, the emission is unaffected because the outflow is quasi-spherical and the characteristic emission time-scales are longer than $R_*/c$. Therefore, here and in the following subsection, we only provide expressions which describe the light curve up until $t=\min(\twrap,t_{\rm obq})$. The emission at times $t>\min(\twrap,t_{\rm obq})$ is as described in Section~\ref{sec:breakoutclasses}.

In the $\theta_{\rm obs}=0$ case, we find
\begin{equation}
    L(t) \approx \begin{cases}
    2\Lbo(c\tbo/R_*), & t<t_{\rm lc} \\
    2\Lbo(c\tbo/R_*) (t/t_{\rm lc})^{(1-k)/(1+k)}, & t_{\rm lc}<t<\min(\twrap,t_{\rm obq})
    \end{cases}.
\end{equation}

On the other hand, for $\theta_{\rm obs}=\upi/2$,
\begin{equation}
    L(t) \approx \begin{cases}
    2\Lbo(c\tbo/R_*)\theta_{\rm lc}^2, & t<t_{\rm lc} \\
    2\Lbo(c\tbo/R_*)\theta_{\rm lc}^2 (t/t_{\rm lc})^{(2-k)/(1+k)}, & t_{\rm lc}<t<\min(\twrap,t_{\rm obq})
    \end{cases}.
\end{equation}
The initial luminosity here is lower by a factor of $\sim \theta_{\rm lc}^2$ compared to the on--axis case.  One factor of $\theta_{\rm lc}$ appears because the light crossing time of a spherical cap of size $\theta_{\rm lc}$ is longer for an observer at the equator, and the other factor of $\theta_{\rm lc}$ comes about because the projected area of this cap is smaller as seen from the equator.

\subsection{Non-instanteous breakout, unaffected by light travel time ($\epsilon > \epsilon_{\rm crit}$)}
For $\epsilon > \epsilon_{\rm crit}$, light travel time does not affect the light curve; $\theta_{\rm lc}$ and $t_{\rm lc}$ are unimportant.  Therefore, for $\theta_{\rm obs}=0$, the emission is the same as described in Section~\ref{sec:breakoutclasses} at all times.

For $\theta_{\rm obs}=\upi/2$, the early light curve is different from the on--axis case simply because of the different projected area of the emitting surface as seen from the equator (see Section~\ref{sec:lighttravel}).  Taking this into account, we have
\begin{equation}
    L(t) \approx \begin{cases}
    \Lbo \theta_{\rm bo}^3, & t<\tbo \\
    \Lbo \theta_{\rm bo}^3 (t/\tbo)^{(2-k)/(1+k)}, & \tbo < t < \min(\twrap,t_{\rm obq})
    \end{cases}.
\end{equation}

\end{document}